\documentclass[journal,hideappendix]{vgtc}        % final (journal style) without appendices
\onlineid{1642}

\vgtccategory{Research}

\vgtcpapertype{theory/model}

\author{%
  \authororcid{Josiah S.\ Carberry}{0000-0002-1825-0097},
  Ed Grimley, and 
  Martha Stewart
}

\authorfooter{
  \item
  	Josiah Carberry is with Brown University.
  	E-mail: jcarberry@example.com
  \item
  	Ed Grimley is with Grimley Widgets, Inc.
  	E-mail: ed.grimley@example.com.

  \item Martha Stewart is with Martha Stewart Enterprises at Microsoft
  Research.
  	E-mail: martha.stewart@example.com.
}

\graphicspath{{figs/}{figures/}{pictures/}{images/}{./}} % where to search for the images

\usepackage{tabu}                      % only used for the table example
\usepackage{booktabs}                  % only used for the table example
\usepackage{lipsum}                    % used to generate placeholder text
\usepackage{mwe}                       % used to generate placeholder figures

\usepackage{mathptmx}     
\usepackage{amsmath}% use matching math font
\usepackage{multirow}
\usepackage{array}
\usepackage{ragged2e}
\usepackage[english]{babel}
\usepackage{csquotes}
\usepackage{enumitem}
\usepackage[usenames,dvipsnames,svgnames,x11names,table]{xcolor}
\PassOptionsToPackage{table}{xcolor}
\usepackage[normalem]{ulem}
\usepackage{caption}
\usepackage{subcaption}
\usepackage{stfloats}
\usepackage{float}
\newif\ifnotes
\notestrue
\usepackage{tikz}
\usetikzlibrary{shapes}
\usepackage{blindtext}

\definecolor{gray}{RGB}{240,240,240}

\definecolor{bck}{RGB}{206,209,215}

\usepackage[skins]{tcolorbox}
\newtcolorbox{mybox}[2]{
    lower separated=false,
    colback=white,
    colframe={#2},
    colbacktitle={#2},
    coltitle=white,
    before upper=\par\noindent{},
    enhanced,
    attach boxed title to top left={yshift=-0.1in,xshift=0.15in},
    boxed title style={boxrule=0pt,colframe=white,},
    title={\color{white}{#1}}
}

\newcommand*{\rebuttal}[1]{{#1}}
\newcommand{\objectives}[0]{\texttt{objectives}}
\newcommand{\intents}[0]{\texttt{intents}}
\newcommand{\strategies}[0]{\texttt{strategies}}
\newcommand{\techniques}[0]{\texttt{techniques}}

\newtcolorbox{definitions}{colback=Thistle!15!white,colframe=Plum!50, , left=0pt,right=0pt,top=0pt,bottom=0pt}

\newcommand{\techname}[1]{\colorbox{gray}{\small \texttt{\color{black}#1}}}

\title{A Multi-Level Task Framework for Event Sequence Analysis}
\author{%
  \authororcid{Kazi Tasnim Zinat}{0000-0001-7914-5955},
  Saimadhav Naga Sakhamuri, Aaron Sun Chen and 
  \authororcid{Zhicheng Liu}{0000-0002-1015-2759}
}

\authorfooter{
  \item
  	Kazi Tasnim Zinat, Saimadhav Naga Sakhamuri, Aaron Sun Chen, and Zhicheng Liu are with the University of Maryland. E-mail: \{kzintas, leozcliu\}@umd.edu, \{ssakhamu, achen151\}@terpmail.umd.edu.
}

\abstract{%
  Despite the development of numerous visual analytics tools for event sequence data across various domains, including but not limited to healthcare, digital marketing, and user behavior analysis, comparing these domain-specific investigations and transferring the results to new datasets and problem areas remain challenging. Task abstractions can help us go beyond domain-specific details, but existing visualization task abstractions are insufficient for event sequence visual analytics because they primarily focus on  \rebuttal{multivariate datasets} and often overlook automated analytical techniques. To address this gap, we propose a domain-agnostic multi-level task framework for event sequence analytics, derived from an analysis of $58$ papers that present event sequence visualization systems. Our framework consists of four levels: \texttt{objective, intent, strategy, and technique}. Overall \texttt{objectives} identify the main goals of analysis. \texttt{Intents} comprises five high-level approaches adopted at each analysis step: augment data, simplify data, configure data, configure visualization, and \rebuttal{manage} provenance. Each \texttt{intent} is accomplished through a number of strategies, for instance, data simplification can be achieved through aggregation, summarization, or segmentation. Finally, each \texttt{strategy} can be implemented by a set of \texttt{techniques} depending on the input and output components. We further show that each technique can be expressed through a quartet of \textit{action-input-output-criteria}. We demonstrate the framework’s descriptive power through case studies and discuss its similarities and differences with previous event sequence task taxonomies.
}


\begin{document}
\maketitle
\section{Introduction}

% \paragraph{1. Importance of event sequence data, how it is ubiquitous in so many domains}

% Event sequence data is ubiquitous across a wide range of application domains. It comprises temporally ordered discrete events for entities in the domain under consideration. 
% In the realm of education, online learning platforms meticulously track student activities, such as video views, assignment submissions, and forum interactions, to gain insights into learner behavior and optimize course content. Professional sports teams collect extensive data on athlete training regimens and in-game performance, leveraging it to enhance strategies and identify areas for improvement. Cybersecurity analysts look into detailed network activity logs, seeking patterns indicative of potential intrusions or malicious behavior. 
Event sequence data contains a wealth of knowledge, but because of their complexity, it can be challenging for analysts to extract useful insights. 
% \paragraph{2. Variety of event sequence tools: Latching on to the domain idea, there are many event sequence analysis tools that have been developed for specific application domains. Provide examples: Tipovis (Social behaviour), Coreflow, SessionViewer (Clickstream), careflow, decisionflow (healthcare) etc. Also mention these tools are trying to solve different challenges, such as comparison, exploration, commonality identification etc}
Numerous visual analytics systems for event sequence data have thus been developed, each tailoring to particular application domains. 
% have been developed to facilitate the understanding of event sequence data.
% The research has led to the creation of numerous customized analysis tools tailored to particular application areas. 
For instance, Tipovis \cite{tiphan2015tipovis} assists developmental psychology researchers in the visual analysis of  
% temporal proximity and overlap of behaviors 
children's behavioral patterns. 
 In clickstream analysis, tools such as Patterns and Sequences \cite{liu2016patternsandsequences}, Coreflow \cite{liu2017coreflow}, and Segmentifier \cite{dextras2019segmentifier} enable market analysts to explore and understand customer journeys. SessionViewer \cite{lam2007sessionviewer} helps analysts examine web session logs. Careflow \cite{perer2013careflow}, DecisionFlow \cite{gotz2014decisionflow}, OutFlow \cite{Wongsuphasawat2011Outflow, wongsuphasawat2012exploring}, and Cadence \cite{gotz2019cadence} help healthcare researchers  in identifying trends  in patient treatment histories and support clinical decision-making. 

% \leo{Emphasize that these tools differ in terms of both the objectives/tasks to be addressed and the techniques used. Give examples and provide references} 
% \zinat{Addressed. Added specific examples and references}
% These tools not only differ in terms of application domains, but also in the objectives they address and the techniques they employ. 
These tools vary in application domains, objectives, and techniques.
Some tools focus on sequence comparison \cite{zhao2015matrixwave, tiphan2015tipovis, qi2020stbins}, while others emphasize overview generation \cite{wongsuphasawat2011lifeflow, wang2008aligning}, pattern exploration \cite{perer2014frequence, liu2017coreflow} or finding similarities \cite{wang2009temporal, monroe2013eventflow}  across sequences. 
% To achieve their respective objectives, 
These tools utilize various techniques, such as sequence alignment \cite{di2020sequencebraiding, zhao2015matrixwave}, pattern mining \cite{liu2017coreflow, liu2016patternsandsequences, vrotsou2018eloquence}, rule-based exploration \cite{cappers2018eventpad2} and novel visual encodings \cite{vrotsou2009activitree}.  Although these domain-specific investigations sometimes hint at the possibility of wider applicability (e.g., \cite{gotz2019cadence, liu2017coreflow}), 
% \leo{need more references here}\zinat{trying to find more papers that mention being adaptable to other domains}, 
the diversity of objectives and techniques 
 % \hannah{diversity in what?} 
 % draws attention to the need for 
 presents challenges in generalizing the 
 % lessons from these works, 
 findings, comparing tools developed for different applications, and transferring knowledge across domains. 
 % Many of these  techniques can be adapted to solve problems in other domains. 

% \paragraph{3. Address the Gap: In many event sequence papers, it is mention, even emphasized that even though the tool has been employed for a specific domain, it can be extended to other domains as well. However, there is no guideline on how to that} \leo{Talk about about why a task framework can close the gap, quote from papers e.g., Sequence Braiding}
% \zinat{Not sure about the order of 4 and 5}

% \leo{we don't usually quote from a specific paper in the intro. The intro should summarize the main ideas, not provide low-level details like this.} 
% \zinat{removed comment}

 % Moreover, despite the domain-specific nature of many event sequence analysis tools, 

 % abstracting analytic intents and identifying common goals across event sequence analysis tools.
 % a comprehensive understanding of the problems that event sequence analysis tools address and the corresponding solutions.
 The development of task frameworks has consistently been instrumental in advancing the field of data visualization, 
 % providing a structured approach to understanding and categorizing user interaction tasks. 
however,  
 existing
 % visualization task 
 frameworks \cite{brehmer2013typology,amar2005lowlevleanalytic,yi2007toward, valiati2006taxonomy, shneiderman2003eyes} 
 fall short of fully capturing the unique challenges associated with event sequence data. 
 % \hannah{are there examples of some of these generic task frameworks you could discuss or cite?} \zinat{cited}
 % Unfortunately, comparable research that focuses on event sequence analysis  in particular  has been missing.
  First, most of these frameworks assume a multivariate data model, 
  % and do not account for 
  overlooking the complexities of event sequence data \cite{di2020sequencebraiding},
such as high dimensionality and time variance \cite{kim2020seq2vec}.  
% For example, groupings in event sequence data can be based on sequences and change over time, unlike the static groupings in tabular data. Summarization and segmentation tasks for event sequence data requires maintaining the chronological order of events. Predictive modeling and what-if scenarios in event sequence data also require temporal awareness. % when a user performs filtering on a multivariate dataset, it is obvious that she is filtering data items (rows) based on attribute (column) values. Filtering in event sequence analytics, however, can have different connotations: filtering sequences (based on events, sequence attributes, or segments), filtering events (keeping the sequences but discarding occurrences of particular events within each sequence) by category or time, or filtering segments (discarding segments but keeping the sequences). 
% \leo{to be revised} \zinat{added my version}
Second, 
% existing frameworks 
they focus mostly on visualization tasks, but event sequence analytics employ visualization in conjunction to pattern mining, unsupervised learning, and data transformation. We need an integrative task framework accounting for all types of techniques. To date, only a few attempts have been made to describe the task space of event sequence visual analytics \cite{du2016volumeandvariety,plaisant2004challenge,peiris2022data}. However, they tend to mix task descriptions of different levels of granularity in a single taxonomy and do not fully capture the complexities of event sequence analysis tasks.  \looseness -1

Inspired by previous works \cite{brehmer2013typology,rind_task_2016} highlighting the multi-dimensional and multi-level nature of visualization tasks, we present a comprehensive end-to-end task framework for event sequence analysis, capturing the analysis process from data preprocessing to provenance tasks. Derived from an extensive analysis of $58$ papers, 
% is created especially for event sequence analysis to overcome the challenges mentioned above. O
our framework consists of four levels  (objective, intent, strategy, and technique),   each capturing a different level of abstraction. 

The highest level, \objectives, represents overarching goals
% that users aim to achieve through their analysis, 
such as cohort comparison, anomaly detection, and identifying common behavior. To achieve these \objectives, users form specific \intents \  at each analysis step. We identify five high-level \intents: augment data, simplify data, configure data, configure visualization, and \rebuttal{manage provenance.} Each intent is realized through a set of \strategies, which define the methods used to accomplish the intent. For instance, data simplification can be achieved through aggregation, summarization, or segmentation strategies. 
Finally, \techniques \ are specific implementations of each strategy, 
expressed as a quartet of dimensions: the \textit{action} performed, \textit{input} data components, desired \textit{output} components, and \textit{criteria} specifying the parameters or conditions for the action.
% and is expressed as a quartet of dimensions \texttt{action-input-output-criteria}: 
% taking into account four dimensions including 
% the \textit{action}, the \textit{input} data components, the desired \textit{output} components, and the \textit{criteria}. 
% \leo{to revise} \zinat{revised}
% We further break down  \texttt{technique} by expressing it with a quartet of \texttt{action-input-output-criteria}.
By organizing tasks into this hierarchical structure with multidimensional characterization at the technique level, our framework bridges the gap between high-level objectives and low-level techniques, providing a comprehensive and systematic approach to event sequence analysis. \looseness -1

We evaluate the expressiveness and precision 
% \leo{need a better word} \zinat{specifity?} 
of our framework by comparing it with existing task abstractions for event sequences \cite{peiris2022data, du2016volumeandvariety, plaisant2016diversity} through case studies, and discuss its implications for future research on event sequence visual analytics. 

\section{Related Work}

% \zinat{provide high-level indication what this section is about}
We first review task taxonomies for interactive visualizations and their limitations in describing event sequence visual analysis.  Then we discuss existing works that discuss tasks in event sequence data.

% \paragraph{Prior work has explored various tasks and operations for analyzing event sequence data. However, there is currently no clear unified theoretical framework that systematically relates the properties of event sequence datasets to the selection of appropriate analysis tasks and workflows. This research aims to address this gap.}

% \paragraph{Furthermore, existing taxonomies of event sequence analysis tasks have some limitations:
% They often mix multiple levels of task specification, lacking a clear hierarchy.
% They tend to focus on only one level of the task abstraction.
% There is a need for a multi-level taxonomy that relates low-level user tasks to higher-level analytic goals.}

\subsection{Task Abstractions for Visualizations}

% \paragraph{Mention Munzner's typology paper and david gotz paper. Mention why these are insufficient for event sequence data.}

% \zinat{Would be good to have some examples}
% \hannah{maybe start with a high-level overview of what these taxonomies are trying to achieve before you start discussing the individual taxonomies.}
Numerous theoretical abstractions of visualization tasks have been proposed, including but not limited to: 
Shneiderman's data types by tasks taxonomy \cite{shneiderman2003eyes},
% proposes a taxonomy of information visualizations categorized by  data types  and tasks. 
Amar et al.'s low-level analytic tasks \cite{amar2005lowlevleanalytic}, 
% present a set of primitive analysis task types  for discussing the analytic capabilities of different information visualization systems. 
Valiati et al.'s formulation encompassing analytic, cognitive, and operational tasks to guide evaluation and design of multidimensional data visualizations\cite{valiati2006taxonomy},
% propose a task taxonomy encompassing . 
Schulz et al.'s design space for visualization task \cite{schulz2013design}, Andrienko and Andrienko's  exploratory data analysis tasks \cite{andrienko2006exploratory}, Yi et al.'s categorization of user intents in interactive visualization \cite{yi2007toward}, and Heer and Shneiderman's 
% argue that the interactive aspects of analysis tools are equally important as visual representation, and 
taxonomy of interactive tasks in visual analysis \cite{heer2012interactive}. 
% The taxonomy emphasizes supporting the entire analytic process, from data specification to exploration, and dissemination. Our framework shares the same motivation.
These taxonomies, however, mostly focus on a single level of abstraction, and some of them mix low-level actions with high-level goals.
\looseness -1
% For example, to \textit{Find Extremum}, the analyst might need to \textit{Sort} the data. However, both of these are mentioned as same level task.

% \leo{we didn't cite Yi et. al.} \zinat{ cited Yi et.al}

% Yi et al. \cite{yi2007interaction} describe interaction techniques based on user intents, and remark low-level categorizations are often inadequate to capture the diversity of interaction techniques.

Researchers have argued that tasks are complex theoretical constructs and cannot be fully understood along a single dimension or at a single level of abstraction. Rind et al. propose the idea of task cube to capture the multi-dimensional nature of visualization tasks \cite{rind_task_2016}. Gotz and Zhou \cite{gotz2008insightprovenance} present a hierarchy of tasks, sub-tasks, actions, and events to characterize user behavior in visual analytics at multiple degrees of semantic granularity. Brehmer and Munzner \cite{brehmer2013typology} introduce a multi-level typology for abstract visualization tasks. The typology is organized around three key questions: why the task is performed, how it is executed, and what the task inputs and outputs are. By utilizing several levels of abstraction, the typology facilitates a more systematic analysis and comparison of visualization tasks across tools and domains. Building upon Brehmer and Munzner's work, Lam et al. \cite{lam2017goalstotasks} derive a framework based on the analysis of design study papers for mapping high-level analysis goals and low-level visualization tasks. Our task framework is inspired by these works' emphasis on the multi-granularity and multi-dimensionality of tasks. 

\rebuttal{These task abstraction frameworks, designed primarily for multivariate data visualization and analysis, are not readily applicable to event sequence visual analytics due to two main reasons. First, event sequence analysis possess unique complexities,
% compared to \rebuttal{multivariate data analysis without sequential aspect \cite{shneiderman2016eventquartet}}. In event sequence data, 
including temporal ordering of events within sequences, and sequences having their own attributes. Additionally, sequences can be grouped into cohorts or divided into segments based on various criteria. These hierarchical complexities and temporal relationships are not addressed in multivariate visualization frameworks \cite{shneiderman2016eventquartet}.  Consequently, event sequence analysis requires a broader set of techniques and a closer examination of the \textit{what} aspects of user actions. For instance, filtering in multivariate datasets is unambiguously about filtering data items based on attribute values, whereas in event sequence analysis, filtering can target various components (e.g., filtering events by  frequency, filtering sequences based on milestone events, filtering sequences by sequence attribute)}
% in the event sequence data model entail two inadequacies of multivariate visualization task abstractions: 1) a much broader set of techniques is needed to analyze event sequences, and 2) a closer examination is needed to address the \textit{what} aspects of user actions. For example, filtering on multivariate dataset implies filtering data items (rows) based on attribute (column) values. In event sequence analysis, however, filtering can be done on a variety of data components.  For example, filtereing events according to their frequency of occurrence as first event within sequences, or filtering sequences based on the occurrence of a series of milestone events as a subsequence.}

 % separating events into sequences according to how frequently they occur as the initial event in each sequence, or grouping sequences together according to whether a sequence of significant events occurred in the sequences
% Task abstractions designed for tabular or graph data, which usually presume a flatter data format, are not able to fully capture these complex relationships and hierarchical structures.

Second, previous works on task abstraction primarily focus on visualization tasks and do not adequately cover the wide range of machine learning and data transformation tasks that are crucial in event sequence analytics.
% \cite{heer2012interactive}. 
For instance, pattern mining techniques are frequently employed to discover recurring event subsequences, and clustering algorithms are used to group similar sequences together. Data transformation tasks, such as attribute or value based filtering, sequence segmentation, and event aggregation, are also common in event sequence analysis. However, existing task frameworks often emphasize visualization tasks, such as looking up data items or comparing attributes, and do not provide a comprehensive taxonomy for data transformation tasks. A comprehensive task framework for event sequence analysis must take into consideration both data manipulation and visualization tasks, and how they are integrated in the end-to-end analysis process. 

\subsection{Surveys and Taxonomies for Event Sequence Visual Analysis}

% Jentner et. al \cite{jentner2019techniquesforpatterns} surveys a wide range of visualization approaches that have been proposed in the literature to represent and explore sequential patterns, and compares their strengths and weaknesses in terms of factors like scalability, pattern identifiability, and ability to convey interestingness measures. Liu et al \cite{liu2016miningpruning} proposes a three-stage framework  for visual analysis of frequent patterns in event sequences. \zinat{What else to mention about this} \leo{I don't think these are task abstractions, we can remove this paragraph.}

% \paragraph{Shneiderman paper, no hierarchy given}

We identify two pieces of work on high-level tasks in event sequence analysis. Plaisant et. al \cite{plaisant2016diversity} propose a characterization of high-level user tasks for event analytics, such as  `Identify a set of records of interest' and `Compare two or more sets of records'. 
% \leo{such as?} \zinat{addressed},  
They also stress the importance of task descriptions for event sequence analytics. This work sets the stage for developing a common language for comparing tools, and applications of event analytics. However, it does not provide a mapping of high-level goals to low-level techniques. 

Guo et al. \cite{guo2021survey} provide a comprehensive survey of visual analytics tools for event sequence data.
% They propose a design space for characterizing event sequence visualizations along four dimensions.
% : Data scale, Analysis method, Visual representation, Interaction technique.
% Using this design space, t
They review and categorize the tools by five high-level analytical tasks (summarization, prediction \& recommendation, anomaly detection, comparison, and causality analysis), 
% h(\leo{list them}  \zinat{listed}), 
and application domains targeted. They further identify seven common interaction techniques for event sequence visual analytics (e.g., filter, segmentation). Our work shares the motivation of Guo et al. to provide a systematic framework for understanding event sequence analysis. While Guo et al. approach this by categorizing tool functionalities 
% design \leo{I don't think it's limited to vis design} \zinat{I agree, should the term be visual task?}
according to high-level analytical tasks,
% \leo{high-level, low-level, or both?} \zinat{addressed},
we instead propose a hierarchical typology that maps analytic intents to specific task strategies and techniques. \looseness -1
% : summarization, prediction \& recommendation, anomaly detection, comparison, and causality analysis.

% \paragraph{Mention about strategies in volume and variety paper, how they are all covering some overarching broader strategy group, but that does not necessarily provide abstractions. Also they mention data manipulation only, nothing related to task.}
For lower-level tasks, Du et. al \cite{du2016volumeandvariety} presents $15$ strategies 
% that analysts can use 
to reduce the volume and variety of temporal event sequence data. While these strategies are a valuable tool set of data manipulation tactics, they do not offer task abstractions that link user goals to system actions, nor do they examine tasks aimed at objectives other than reducing volume and variety (e.g., augmenting event sequences for analysis).
% These strategies also vary in terms of level of abstraction, \leo{for example, ...}\zinat{they all map to low level tasks}. 
% Nevertheless, these common actions in the analysis process provide a valuable starting point for the development of a comprehensive task abstraction. 
% A more complete task abstraction for event sequence analysis would need to consider not only data manipulation but also task-specific strategies, and intents behind the actions.
% \hannah{the transition between paragraphs in this subsection is awkward. Try to connect the works you're covering to each other.}

% However, these papers focus on

% \paragraph{Shunan's paper: surveyed and categorized papers but no hierarchy}
% \subsection{Previous Survey and Task for Event Sequence Data}

% Furthermore, our framework puts greater emphasis on data manipulation tasks in addition to visual representation and interaction.

 % \paragraph{Workshop paper}: Most similar to our work. presents a methodology and task typology for time-stamped event sequences (TSES) that has similarities to our current work. Both share motivations to develop taxonomic structures that characterize analysis tasks for sequential temporal data in a systematic, data-centric way. Additionally, the methodology follows Munzner's nested model by mapping low-level user tasks to high-level analytic goals.

 A recent workshop paper by Peiris et al.  \cite{peiris2022data} presents a methodology and task typology for time-stamped event sequences (TSES) that has similarities to our work. They provide a list of $23$ tasks described in terms of the action, target, and criteria triangle. 
 The paper has a narrower focus on TSES which has continuous values, whereas our work considers a broader scope including both TSES and event sequences with ordered events. These strategies also vary in terms of level of abstraction. 
 % , for example, `Analyze Trends', `Show Details'. 
 In addition, while the paper addresses the multi-dimensional nature of tasks, it does not provide an account of tasks across multiple levels of granularity. 
 % and focuses on building a connection between high-level goals and low-level techniques. 
 % Both works share motivations to develop abstractions of analysis tasks for sequential temporal data in a systematic, data-centric way. however, the methodology in \cite{peiris2022data} follows Munzner's nested model by mapping low-level user tasks to high-level analytic goals, and does not provide hierarchical levels.
In section \ref{sec:eval}, we provide a comparison of tasks from our framework with the tasks proposed by these three works \cite{plaisant2016diversity,peiris2022data,du2016volumeandvariety}. 

\section{Method}
We adopt an iterative approach to develop the task framework based on empirical data from literature review.  Literature reviews are a widely adopted method for task abstraction \cite{gotz2008insightprovenance, yi2007toward, ahn2013task}.  
% To tackle the validity threats\cite{kerracher2017constructing}, 
% we utilize the power of Large Language Models to understand domain specific terms \zinat{cite LLMs??}, 
% \leo{I would remove this} \zinat{I am inclined to keep this} we review the methodology of previous papers that perform domain agnostic task generation \cite{lam2017goalstotasks, wang2022ml4vis}. \hannah{How does this tackle the validity threats? Also what type of validity?}
Our process consists of two main stages: open coding and axial coding. 

\subsection{Corpus Assembly}
% \zinat{
%     1. Scope of the search space (e.g., relevant conferences, journals, and time period) - Top HCI and Vis conferences, such as CHI, IEEE Vis, EuroVis. Also looked into previous survey papers to add in more event sequence paper
%     2. inclusion criteria for papers (e.g., must analyze event sequence data with categorical information, we exclude papers analyzing Time stamped event sequences (TSES), and time series data) 
%     3. Conducting a first pass to identify papers that meet the initial criteria
%     % 4. \textcolor{red}{Apply additional criteria to refine the paper set - what would be additional criteria for us? }
%     }
% \zinat{Remove thorough based on reviewer comment? Rewrite?} 
    
\rebuttal{We conducted web searches to curate papers presenting event sequence systems published at HCI and visualization venues. A complete list of keywords is in the supplemental materials  \footnote{All the supplemental materials are made available as \href{https://osf.io/bkjsc/?view_only=b95871b8c4ae497ab9b6cb565e28edf5}{an OSF project}}.
 % using keywords related to event sequence analysis. Details of venues and keywords are in the appendix. 
Additionally, we examined references from Peiris et al. \cite{peiris2022data} and Guo et al. \cite{guo2021survey} to ensure comprehensive coverage. We conducted a first pass through titles, abstracts, and keywords of papers from selected sources, identifying $105$ papers that potentially met our criteria for analyzing sequential data.}

    % We conducted a thorough literature review of event sequence systems published at top HCI and visualization conferences and journals, searching for keywords related to event sequence analysis. \rebuttal{ Refer to the appendix for the details of the venues and keywords.}
    % % CHI, IEEE VIS, EuroVis and TVCG, 
    % % searching for keywords,  `event', `sequence', `sequential', `progression', `temporal', `pattern', \rebuttal{`stage', `time-stamped'}. 
    % % \leo{what search keywords were used?} \zinat{answered}. 
    % We also collected papers by examining the references of the methodology and task typology paper by Peiris et al. \cite{peiris2022data} and the survey paper on event sequence analysis  by Guo et al \cite{guo2021survey} to expand our coverage and include important works that may have been published in other venues. To identify papers that met our initial criteria of analyzing sequential data,  
    % % \leo{what is this initial criteria?} \zinat{contains these keywrds}, 
    % we conducted a first pass through the titles, abstracts, and keywords of papers from the selected conferences and survey papers. During this pass, we marked papers that potentially met the inclusion criteria for further review, resulting in $105$ papers. 
    % \zinat{Mention initial number of papers}

\rebuttal{After the first pass, we refined our selection by including papers that analyze event sequence data with at least one categorical attribute, excluding those with only numerical attributes \cite{lekschas2020peax}
% \hannah{does this mean at least 1 categorical attribute or only categorical attributes?}. 
% We excluded papers with events that have numerical attributes only \cite{lekschas2020peax} 
% \leo{I do think we cover time-stamped event sequences. We just don't cover events that have numerical attributes only.} 
% \zinat{Fixed} 
or time series data \cite{bernard2012guided}, as these types of datasets have different characteristics and analysis requirements compared to categorical event sequences. We read the full text of the marked papers to confirm they met our inclusion criteria, resulting in $58$ papers from $16$ venues. The supplemental materials include detailed information on the distribution of papers across venues and years.  
% \leo{Was the criterion applied in this initial pass?} \zinat{We applied it in the second pass, when going through the papers} \zinat{resulting in X number of papers}
We also noticed that some systems were covered by multiple papers, such as EventAction \cite{du2016eventaction, du2019eventaction}, LifeLines2 \cite{wang2009temporal, wang2008aligning}. Thus, our final list contained $52$ systems.}\looseness -2
 % \hannah{I've seen tables in literature reviews that cover the different papers that were surveyed and the themes found across each of them. I think that might be beneficial to this work. Maybe you could add that to~\autoref{tbl:quartet}}\zinat{Not inclined to add, since the papers reviewed by undergrads need a intensive second pass}

% \zinat{ Papers that did not provide sufficient detail on the tasks and techniques used for event sequence analysis were excluded.} \leo{mention one or two example papers that were excluded}\zinat{I cannot recall if we excluded any,  should delete this paragraph}

\subsection{Open Coding}
% \zinat{
%     1. Assigned three authors to perform open coding, with a fourth author acting as a tie-breaker
%     2. Identify paper sections containing sufficient information to delineate goals and techniques for event sequence analysis
%     For each relevant section: a) Split it into micro part of sentences/ clauses containing one user or system action (Example: pattern mining based on MDL, double-click the glyphs to expand them for detailed analysis) b) Tag each micro part with the technique c) each micro part maybe associated with multiple technique (Example: Drilling down into a branch to get sub-pattern - contains both filtering and segmenting action)
%     3. Emphasize the iterative and collaborative nature of the opencoding process}

In open coding stage, we first identified paper sections with sufficient information to describe low-level actions. These sections typically included system descriptions. The coders carefully read through each relevant section and split the text into micro parts, consisting of sentences or clauses that described a single user or system action. For example, a micro part could be ``\textit{pattern mining based on MDL}'' or ``d\textit{ouble-click the glyphs to expand them for detailed analysis}'' \cite{chen2017sequencesynopsis}.
\looseness -1
% \leo{are these quotes from the original papers?} \zinat{Yes}

Each micro part was then tagged with one or more labels based on the action(s) described. The coders used a preliminary set of labels derived from their domain knowledge \cite{plaisant2016diversity, du2016volumeandvariety} and expanded the set as new actions emerged from the data. 
In our labeling, we paid special attention to two issues. First, a micro part could comprise more than one action and hence be associated with multiple 
labels. For instance, the micro part ``\textit{drilling down into a branch to get sub-pattern}'' \cite{liu2017coreflow} involves two actions, filtering and segmenting. Second, a verb alone often could not capture the richness and multi-dimensional nature of an action, and we need to include information on the data components associated with each action. For instance, in ``\textit{user can sort page groups alphabetically, by volume, or difference. Sorting can be performed within a specific step or across all steps}'' \cite{zhao2015matrixwave}, 
while the fundamental action remains the same (i.e., sorting), the specific outcome of the action can vary significantly depending on the data elements being sorted (within a specific step or across all steps) and the conditions used to determine the sorting order (alphabetically, by volume, or difference).  \looseness -1

% \leo{provide a simple example here without explicitly naming them as input, output, criteria.}\zinat{added}
% We also took note of the data components associated with each action, such as sequence attributes, event attributes, and other relevant data properties. This information was crucial because a verb alone often could not capture the richness and multi-dimensional nature of an action.
% for understanding the relationship between data and techniques. 
% \leo{We can move this to the previous paragraph to talk about the multidimensional labels.} \zinat{moved.} 
% Augmenting the information about data properties, each of the labels can be  extended to four dimensions of action-input-output-criteria, capturing the full picture of transformation. 

% , and these actions apply on a set of sequences (input) to obtain a subset (output) of sub-patterns (output), based on selected branch (criteria).

% labels. For instance, the micro part `Drilling down into a branch to get sub-pattern' \cite{liu2017coreflow} involves both filtering and segmenting actions. 

% \zinat{added high-level goal details}
\rebuttal{While the bottom-up coding approach provides a detailed description of each analytic step, we recognize that data analysis is an emergent process \cite{emmertstrreib2016emergent}  that cannot be effectively expressed by simply accumulating discrete steps. Therefore, we also considered the holistic objectives of each analysis, leading us to include high-level goals or objectives as a separate level in our framework, 
% complementing the detailed descriptions of the analytic steps.
% By incorporating this additional level, our framework captures not only the intricate details of each step, but also the overarching objectives that guide the entire analytic process. 
% The high-level goals provide a broader perspective on the purpose and desired outcomes of the analysis, while the detailed descriptions of the steps offer insights into the specific techniques and actions employed to achieve those goals. 
This dual perspective ensures our framework comprehensively represents the complex and emergent nature of data analysis in the context of event sequence visualization.}
\looseness -1
% The bottom-up coding approach that we adopted delves into describing each analytic step in details. However, data analysis is an emergent process \cite{emmertstrreib2016emergent} that is not just an accumulation of analysis steps. Therefore, we also note the holistic goal of each analysis process mentioned in the paper, which led to adding high-level goals or analysis objectives as a separate level in our framework that covers the entire analytic process. 

% \leo{we should mention that we did not just focus on assigning the technique names (the verb), we also coded the input/output etc. highlighting that our effort focuses on the multi-dimensional nature of techniques, and we probably should not use the word ``technique'' here, just mention we focus on these different dimensions.}\zinat{addressed}

% To reduce the potential for individual bias in the coding process, three authors independently performed the coding on the paper set, with a fourth author acting as a tie-breaker in case of disagreements or conflicts. 
% % This setup reduced the potential for individual bias in the coding process.
% The open coding process was highly iterative and collaborative. Weekly meetings were held over six months 
% % \hannah{over X months} 
% to discuss and compare the coding results, ensuring consistency and refining the coding scheme as needed. The process was repeated until there were no more disagreements. 

\rebuttal{To reduce potential bias, three authors independently performed coding on the paper set, with a fourth author acting as a tie-breaker in case of disagreements. The open coding process was iterative and collaborative, involving weekly meetings held over six months to discuss and compare results. This ensured consistency and refinement of the coding scheme until all disagreements were resolved.}

 \rebuttal{It is important to note we \textit{focused primarily on actions that result in observable changes in data representation or visual display; leading to tangible transformations, such as the way data is processed, analyzed, or presented}. As such, actions that involve only cognitive processing in humans (e.g., saw, noticed, followed, observed, visually scanned for comparison) without any direct manifestation in the system are outside the scope of this work. This decision ensures the coded actions have a clear and traceable connection to system functionality.} 

% \zinat{interaction techniques}

 \rebuttal{We also omitted low-level interaction details on how an action is accomplished through user interface. For instance, zooming can be implemented using dedicated zoom controls (e.g., buttons, sliders, scale), double-clicking, or through brushing. These low-level features vary significantly across tools and are less critical for understanding high-level tasks and actions performed by users \cite{yi2007toward}.}

% \leo{mention focusing on tasks that have observable changes in data representation or visual display}\zinat{added paragraph}

\subsection{Axial Coding} 
\label{sec:axial}
% \leo{call this subsection axial coding?} \zinat{done}

% \zinat{
%     1. Used affinity diagramming to clarify, merge, split, and add techniques
%     2. Took note of the associated data components with the action, such as sequence attributes, event attributes etc.
%     3. Came up with definitions for the tasks
%     4. Discover relationships and structures among the techniques and create higher-level abstraction Intent  
%     5. Realize the need to  
%     Refine the framework into data and vis parts based on insights gained during affinity diagramming, and created another level Objective.
%     6. Identify High level goals that encmpass the entire system
% }
The open coding resulted in the identification of four primary dimensions for each labeled micro part: \textit{action}, \textit{input}, 
% (data or visual components acted upon), 
\textit{output},
% (generated/modified data or visual components), 
and \textit{criteria}.
% (parameters or conditions associated with the action). 
For example, in `` $\dots$ \textit{incorporate HMMs for exploring disease progression patterns from longitudinal health records}'' \cite{kwon2020dpvis}, the action is \textit{extract}, input is a set of sequences (health records), output is a set of latent patterns, and the criteria is Hiddden Markov Models or HMM. 
% \leo{provide an example with these four dimensions cleared called out.} \zinat{added} 
We then performed axial coding to group these labels into categories spanning three levels: techniques, strategies, and intents.
% \techniques, \strategies, and \intents. 
We used affinity diagramming to organize the identified \textit{action-input-output-criteria} quartets into \techniques. Through this process, we formed clear definitions to ensure consistency, merged similar techniques, and split compound techniques into smaller units. 

\begin{definitions}
\noindent\textbf{Techniques}: each technique is characterized by the following four dimensions:

the \textit{action} performed,

the \textit{input} data or visual components acted upon, 

the generated or modified \textit{output} components, and 

the \textit{criteria} specifying the parameters or conditions associated with the action.     

% (data or visual components acted upon), \texttt{output} (generated/modified data or visual components), and \texttt{criteria} (parameters or conditions associated with the action)

\end{definitions}

% \leo{define technique in terms of the dimensions used in the labels} \zinat{addressed} 

% Based on the refined set of techniques, we developed clear definitions for each task. These definitions established a common understanding of the tasks and their scope, ensuring consistency in the subsequent stages of framework development.

% \begin{definitions}
We organized techniques into groups based on similarities in nature of their \textit{action}, \textit{input} and \textit{output} components, which led to the creation of higher-level abstractions called \strategies.  
% \end{definitions}
For example, `identifying common patterns' and `extracting latent patterns' in the data were grouped under the `Summarize' strategy as they both produce patterns (\textit{output}) from sequences (\textit{input}), and provide a summary of the dataset.

% \begin{definitions}
% \noindent\textbf{Strategies}: \leo{add definition here.}
% \end{definitions}

While \techniques \ and \strategies \  answer  `how' each analysis step is performed, they do not shed light upon `why' these steps are being executed. Therefore, we introduced an additional level of abstraction called \intents \ by grouping related strategies based on the underlying motive.  
For instance, the \strategies \  `Summarize' and `Aggregate' are grouped under the intent \  of `Data Simplification' as they both seek to simplify event sequence data.
% \begin{definitions}
% \noindent\textbf{Intents}: \leo{add definition here.}
% \end{definitions}

% During the coding process, we observed strategies could be categorized based on whether they primarily operated on the data itself or focused on the visual representation of the data. \hannah{explain why this is needed} \todo{still unanswered, cannot find a way to express}. Subsequently, we further refined our framework by separating the `what' component; explicitly distinguishing between data-centric and visualization-centric intents. 

%During the affinity diagramming process, we realized the need to further refine our framework into two distinct components: data and visualization \hannah{explain why this is needed}. This insight prompted us to create another level of abstraction called \texttt{Intents}, which grouped related strategies. For instance, the strategies `Summarize' and `Aggregate' are grouped under the intent of `Data Simplification' as they both simplify event sequence data.
Finally, we identified overarching objectives that encompass the entire analysis process, called \objectives. 
% \begin{definitions}
% \noindent\textbf{Objectives}: \leo{add definition here.}
% \end{definitions}
These \objectives \ guide the design and development of strategies and techniques in the
event sequence analysis tools. 
% and serve as a good indication of strategies and techniques present in the system. 
For instance, earlier tools like LifeFlow \cite{wongsuphasawat2011lifeflow} and Lifeline \cite{plaisant1996lifelines} primarily facilitate pattern exploration, while subsequent tools such as MatrixWave \cite{zhao2015matrixwave} and Coco \cite{malik2015coco} focus on supporting cohort comparison. DPVis \cite{kwon2020dpvis}, on the other hand, specifically aims to model progression pathways. 
% Based on the analysis use cases, o
One system may cover multiple analysis goals, for example, the analysis goals in Sequence Synopsis \cite{chen2017sequencesynopsis} are temporal pattern exploration and correlation analysis.
 % \leo{I don't think these goals are derived in a bottom-up manner from the techniques, these are extracted separately. We should talk about labeling high-level goals in the open coding stage too.} \zinat{addressed}
\section{Framework} \label{sec:  framework}

\begin{figure*}[tb]% specify a combination of t, b, p, or h for top, bottom, on its own page, or here
  \centering % avoid the use of \begin{center}...\end{center} and use \centering instead (more compact)
  \includegraphics[width=\linewidth, alt={Task Abstraction Framework for Event Sequences.}]{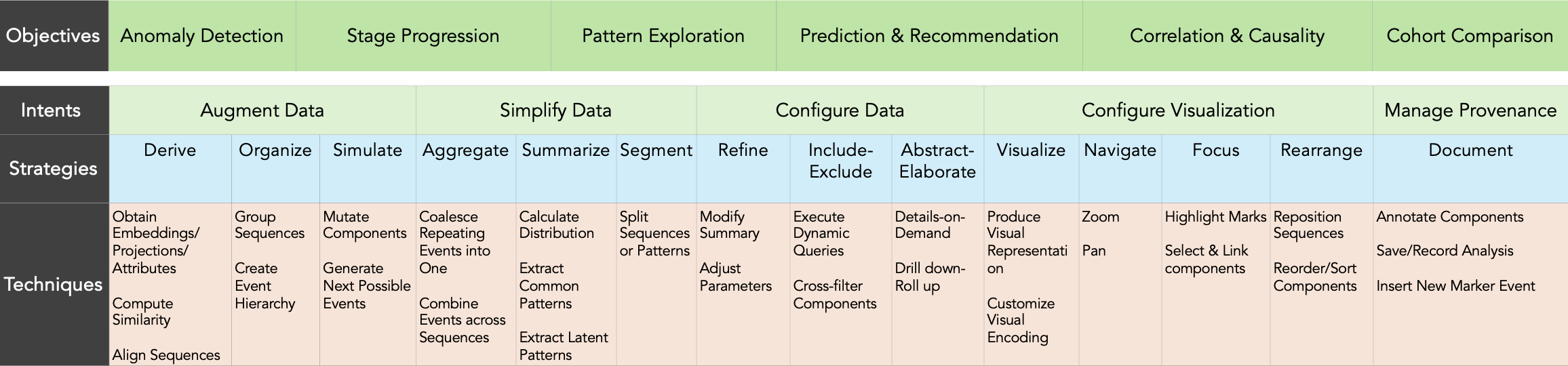}
  \caption{Our multi-level task framework consists of four hierarchical levels: \texttt{objectives}, \texttt{intents}, \texttt{strategies}, and \texttt{techniques}. We identify six overarching \texttt{objectives} of event sequence analysis. Each analysis step is associated with one of the five \texttt{intents}, depicting the purpose of the analysis step. Intents are realized through multiple \texttt{strategies}. Finally, a wide range of \texttt{techniques} are available for implementing each strategy. 
  	% The figure displays Intents, corresponding strategies, and a list of techniques that may be used to implement these strategies \todo{Update Figure with goals. \leo{Consider moving the figure to the first page as the teaser.}}
  }
  \label{fig:strategytechnique}
\end{figure*}

% \zinat{
%     1. Finalize the structure and organization of the event sequence task typology
%     2. Define and describe each level of the typology (e.g., goals, intents, techniques)
%     3. Provide examples for each component of the typology based on the coded analysis reports
% }

% \leo{Provide an overview of the four levels in the opening paragraph in this section. Then talk about high-level goals in the next paragraph.} \zinat{moved the discussion to previous section, axial coding}

Our framework comprises four levels of tasks (\cref{fig:strategytechnique}). At the highest level, the overarching analysis \objectives \  capture six high-level goals that users aim to achieve: anomaly detection, stage progression, pattern exploration, prediction \& recommendation, correlation \& causality analysis, and cohort comparison. 

Each objective involves realizing five main types of analysis \intents: augment data, simplify data, configure data, configure visualization, and manage provenance. 

\rebuttal{Each intent can be accomplished through different \strategies, depending on the data or visualization components involved and the analysis approach. For example, the intent to simplify sequence data can be achieved using aggregation, summarization, or segmentation.} 
 
% through executing a sequence of intents, strategies, and actions at each analysis step. Examples include pattern discovery, cohort comparison, and anomaly detection. Formally, it can be described as $\texttt{high-level goal}:= \texttt{analysis objective} \approx (\texttt{strategy}_{t_1} \rightarrow \dots \rightarrow \texttt{strategy}_{t_i})$.
% , each capturing a different aspect of the event sequence analysis process. 
At the most granular level, the framework defines \techniques, which are the concrete, actionable analysis steps, such as creating new attributes, coalescing repeating events, or modifying summary patterns. Each technique can be described via a quartet of 
\textit{action}, \textit{input}, \textit{output}, and \textit{criteria}, which have been defined in \cref{sec:axial}.
% using the following notation $\texttt{technique} := \textless \texttt{action}, \texttt{input}, \texttt{output}, \texttt{criteria}\textgreater$. 

% Techniques are grouped into strategies, therefore each technique is a set of strategies, such as derive, simulate, segment. $\texttt{strategy} :=\{\texttt{technique}_1,\dots\texttt{technique}_n\}$. 

% A collection of strategies form an intent, representing the purpose of an analysis step, such as data augmentation, simplification, or configuration. $\texttt{intent} :=\{\texttt{strategy}_1,\dots\texttt{strategy}_n\}$. 
% Finally, the high-level goals capture the objective of analysis $\texttt{high-level goal}:= \texttt{analysis objective}$

% At the highest level, the framework identifies the overarching analysis objectives or high-level goals that users aim to achieve through executing a sequence of intents, strategies, and actions at each analysis step. Examples include pattern discovery, cohort comparison, and anomaly detection. Formally, it can be described as $\texttt{high-level goal}:= \texttt{analysis objective} \approx (\texttt{strategy}_{t_1} \rightarrow \dots \rightarrow \texttt{strategy}_{t_i})$.

% During our analysis of the event sequence literature, we observed that each analysis usually has an overarching high-level goal associated with it. These goals often guide the design and development of the event sequence analysis tools and serves as a good indication of strategies and techniques present in the system.  

\subsubsection*{Objectives}
\rebuttal{We identified six common high-level goals providing a broad categorization of \objectives \ encompassing the majority of the papers:}
\begin{itemize}[leftmargin=0pt]
\itemsep0em 
    \item[] \textbf{Anomaly Detection:} Identifying rare  or unexpected patterns or events that deviate significantly from expected behavior. Example: healthcare \cite{Guo2019anomaly} \looseness -1 
    \item[] \textbf{Stage Progression:} \rebuttal{Capturing evolution of processes across distinct stages or phases within sequences, by segmenting sequences and uncovering trends, paths, and factors influencing the transition over time. Example: healthcare \cite{kwon2020dpvis}}
    \item[] \textbf{Pattern Exploration:} Discovering common patterns in event sequence data through exploratory analysis. Example: healthcare \cite{magallanes2021sequen}, 
    % online education \cite{johansson2019moocad}, 
    sports \cite{wu2022rasipam}, clickstream \cite{liu2016patternsandsequences}, manufacturing \cite{j02014livegantt} 
    \item[] \textbf{Prediction \& Recommendation:} Modeling historical event sequence data to forecast future outcomes, and provide recommendations. Example: marketing \cite{guo2019uncertainty}, healthcare \cite{Kwon2018RetainVisVA}
    \item[] \textbf{Correlation \& Causality:} Investigating  dependencies, and potential causal links between events. Example: social media \cite{wu2014opinionflow}, healthcare \cite{jin2021seqcausal} 
    \item[] \textbf{Cohort Comparison:}  Analyzing the event sequences of different subpopulations to identify similarities, differences, and patterns specific to each cohort. Example: healthcare \cite{magallanes2021sequen}, social media \cite{krause2015supporting} \looseness -1
\end{itemize}

% These goals provide a broad categorization of the \objectives \ that serve as analysis goals. 
 % \hannah{should this be it's own subsection?} \leo{Agree, it should have its own subsection, and also these high-level goals should be more described in more detail, with examples. Consider using an itemized list at least. } \zinat{Used itemized list, wrote definitions}

% \zinat{After identifying  overall analysis goal, we associate each analysis step to an underlying intent. We identified five intents: Augment Data, Simplify Data, Configure Data, Configure Visualization, Create Provenance. there are strategies through which these intents can be realized, such as Summarize, agggregate, segment. Finally, based on the input and putput charachteristics, we identify some low-level techniques that serve as the practical implementations of the strategies. Our trchnique list is not exhaustive.}

\subsubsection*{Intents, Strategies, and Techniques}

% Our framework identifies five key intents: Augment Data, Simplify Data, Configure Data, Configure Visualization, and Create Provenance. These intents represent the purpose of the analysis step within the analysis workflow. Each intent can be realized through a set of strategies, which define the general approaches employed to accomplish the intent.
% For example, the Simplify Data intent can be achieved through strategies such as Summarize, Aggregate, or Segment. At the lowest level of the framework, we identify specific techniques that serve as the practical implementations of the strategies, taking into account the input data characteristics and the desired output representations. For instance, the Summarize strategy involves generating concise overviews by computing statistical distributions, and extracting common or latent patterns. 

\rebuttal{We now describe the \intents, \strategies, and \techniques \ in our task framework. \cref{fig:strategytechnique} lists the identified \intents \ and \strategies.  \cref{tbl:quartet} further elaborates on each technique along the \textit{action, input, output,} and \textit{criteria} dimensions. 
It is important to note that the list of \techniques \  is \textbf{not exhaustive} and can be extended based on new systems and case studies.
Depending on system design and implementation, the techniques may or may not involve a visualization display that dynamically updates with user action.} 
% based on the specific requirements of analysis.
% We now describe the intents, strategies, and techniques in our task framework. Figure 1 lists the identified intents and strategies. Table 1 elaborates on each technique along the dimensions of action, input, output, and criteria. This list is not exhaustive and can be extended with new systems and case studies. Depending on system design, these techniques may or may not involve a dynamically updating visualization display.

% \afterpage{%

\begin{table*}[p]\fontfamily{ptm}\selectfont\small
\caption{The \textit{action-input-output-criteria} quartet characterizing \texttt{techniques} in our multi-level task framework. Each technique is defined by the \textit{action} performed, the \textit{input} data components, the desired \textit{output} components, and the \textit{criteria} guiding the transformation.}
\label{tbl:quartet}

\centering
{\small
\begin{tabular}{p{0.07\linewidth}p{0.2\linewidth}p{0.2\linewidth}p{0.43\linewidth}}
\hline
\textbf{Action} & \textbf{Input} & \textbf{Output} & \textbf{Criteria} \\
\hline
Obtain & Event Sequence Data & Embeddings/ Projections/ Attributes & Statistical or machine learning algorithms, such as TF-IDF, dimensionality reduction technique (e.g., t-SNE)
% , event or sequence embedding 
\\
\hline
Compute & Events, Sequences, Patterns & Similarity & Statistical or machine learning algorithms, such as distance metric (e.g., Jaccard Index, Levenshtein Distance)  \\
\hline
Align & Set of Sequences & Aligned Sequences & Common reference point (e.g., origin time), specialized alignment procedures, (e.g., proxy event insertion, Dynamic Time Warping, shortest common supersequence) \\
\hline
\multirow{6}{*}{Group} & \multirow{5}{*}{Set of Sequences} & \multirow{5}{*}{Grouped Sequences} & Event: entry event, focal event of interest \\
\cline{4-4}
&  &  & Event attribute: category, event attribute ranges \\
\cline{4-4}
&  &  & Sequence attribute: categorical \& continuous \\
\cline{4-4}
&  &  & Derived: clustering, similarity to prototype \\
\cline{4-4}
&  &  & Pattern: event subsequence (milestone) \\
\cline{2-4}
& Set of Events & Event Hierarchy & Automated algorithms, existing hierarchical structures, regular expression or user-defined criteria \\
\hline
Mutate & Event, Event Attribute, Patterns, Sequence Attribute & Mutated sequences & User-defined scenario (e.g., altered event sentiment) \\
\hline
Generate & Sequence & Next possible event & Machine learning model (e.g., trained on historical event sequences) \\
\hline
Coalesce & Sequence & Simplified Sequence & Repeating events of same type, frequently occurring event bundle \\
\hline
Combine & Set of Sequences & Tree/DAG & Event and ancestor events \\
\hline
Calculate & Set of Sequences, Events, Event Attributes, Sequence Attributes & Statistical Summary (e.g., histograms, charts) & Statistical formulae \\
\hline
Extract & Set of Sequences & Common Patterns & Mining algorithms (e.g., VMSP, Frequent Sequential Pattern, PrefixSpan, Multiple Sequence Alignment) \\
\cline{3-4}
&  & Latent Patterns & Statistical or machine learning algorithms (e.g., HMM, CP Decomposition, hierarchical Bayesian models) \\
\hline
Split & Sequence, Pattern & Sequence or Pattern Segments & Event (e.g., landmark events), pattern (e.g., milestones), derived latent stages (e.g., content vector segmentation) \\
\cline{4-4}
&  &  & Time: fixed time intervals (e.g., year, month, day), \\
\hline
Modify & Summary Patterns & Modified Summary Patterns & User domain knowledge (e.g., add/delete/edit events, merge/split patterns), thresholds (e.g. minimum support, gap tolerances, number of hidden states) \\
\hline
Adjust & Current data processing and modelling specifications & Updated data processing and modelling specifications  & Setting value to parameters, (e.g., window size, similarity threshold, importance measure, constraint/filter criteria) \\
\hline
\multirow{4}{*}{Query} & \multirow{4}{=}{Set of Sequences, Patterns, Events, Event Attributes, Sequence Attributes} & \multirow{4}{=}{Query Output (filtered data)} & Event Attribute: categorical event types, event attribute ranges, event occurrence time, intervals, duration, derived event attribute \\
\cline{4-4}
&  &  & Sequence Attribute: categorical and continuous sequence attributes, derived sequence properties (e.g., total duration, event count) \\
\cline{4-4}
&  &  & Pattern: subsequences, time intervals, keywords, transitions, presence/absence of event set \\
\cline{4-4}
&  &  & Advanced querying techniques: graphical queries, regular expressions \\
\hline
Cross-Filter & Data (e.g., Events, Event Attributes, Sequence, Sequence Attributes, Patterns) glyph in one view & Filtered Data in coordinated views & Data attributes or dimension selection in other views \\
\hline
Display & Data (e.g., event, sequence, pattern) glyph & Detailed Information & Selected data component \\
\hline
Drill down - Roll up & Hierarchical data organization & Detailed View (Drill down) or Summary View (Roll up) & User selected granularity \\
\hline
Produce & Event Sequence Data & Visual Representation (e.g., timeline, graph, aggregate view, multivariate view) & Visual encoding rules (e.g., position, size, color, shape, opacity) \\
\hline
Customize & Visual Representation & Customized Visual Representation & User preferences (e.g., color palette, node size/shape, edge style, glyph design, visual variable mapping) \\
\hline
Update & Layout & Updated layout & System-specific control(e.g., comparison mode toggle, update canvas size) \\
\hline
Zoom & View & Zoomed View & Zoom level \\
\hline
Pan & View & Panned View & Pan direction \\
\hline
Highlight & Marks (e.g., events, patterns, sequences) & visually emphasized mark (e.g., change in color, opacity, border, glyph) & User selection \\
\hline
Select & Data (e.g., Event, Event Attributes, Sequence, Sequence Attributes, Patterns) glyph & Linked Highlighting & User selection and underlying data corrrespondence \\
\hline
Reposition & Set of Sequences & Repositioned Sequences & Alignment event \\
\hline
Reorder/ Sort & Events, Event Attributes, Sequence, Sequence Attributes, Patterns & Reordered/Sorted Components & Sorting criteria (e.g., frequency, alphabetical, temporal, correlation, prediction accuracy, attribute values, similarity, custom order) \\
\hline
Annotate & Data glyph & Annotated Representation & User notes, comments, descriptions \\
\hline
Save/ Record & Analysis State & Saved/Recorded Analysis & User-specified snapshot \\
\hline
Insert & Sequence & Sequence with Marker Event & User-defined timestamp \\
\hline
\end{tabular}
}
\end{table*}

% \zinat{Should we include the paper name in example, or should we just use the reference?} \leo{paper name should be included}

\subsection{Augment Data}
Data Augmentation refers to enhancing raw event sequence dataset by adding, modifying, or deriving new components, attributes, 
representations, or relationships that were not initially present in the dataset.

 % adding/modifying attributes/representations/relationships that were not present in the raw dataset: derive, group, simulate
\subsubsection{DERIVE} \label{derive}
% \leo{we need some visual cue to differentiate the intent from strategy from technique, maybe using some background color or some font variations. }

Derive entails conducting diverse operations on event sequence data to compute additional attributes or representations necessary for downstream analysis. In the data preprocessing pipeline, multiple techniques can be sequentially employed  to achieve this strategy.
% It involves applying data transformation, computing event attributes, event or sequence similarity and using sequence alignment techniques to enhance the usability. 

% \paragraph{Expand Attribute Set}
% Computing Event Attribute features 
% refers to expanding the attribute set by calculating additional attributes or properties. It 
% involves applying statistical or machine learning techniques to extract meaningful features as new event attributes or sequence attributes.
% enriching  events, sequences with supplementary information. 
% For example, Sentiment-Specific Word Embedding (SSWE) is used to capture sentiment-related contextual information in OpinionFlow\cite{wu2014opinionflow}, ProtoSteer \cite{ming2019protosteer} calculates event embedding, event similarity, and event alignment sequence embedding, and EventThread \cite{guo2017eventthread} calculates TF-IDF.
% \leo{which of these are event attributes and which are sequence attributes?} \zinat{SSWE, TF-IDF- event attribute, event embedding - event attribute, event similarity - shared between two events, and event alignment sequence embedding- sequence attribute}

\noindent \techname{Obtain Embeddings/Projections/Attributes:}
transforming raw event sequence data into a more structured, and analysis-ready format via augmenting embeddings or projections to facilitate subsequent modeling, and analysis. This often involves applying statistical or machine learning methods. For example EventThread \cite{guo2017eventthread} calculates TF-IDF and creates a three-way tensor representation of sequence data. 
% Similarly, in EventThread2 \cite{guo2018eventthread2}, events are mapped to embeddings to capture the co-occurrence likelihood of events and create a vector representation. 
ProtoSteer \cite{ming2019protosteer} calculates event and sequence embedding, and  employs t-SNE to project the high-dimensional sequence embedding into a lower-dimensional space.
% For example, Sentiment-Specific Word Embedding (SSWE) is used to capture sentiment-related contextual information in OpinionFlow\cite{wu2014opinionflow}, ProtoSteer \cite{ming2019protosteer} calculates event embedding, event similarity, and event alignment sequence embedding, and EventThread \cite{guo2017eventthread} calculates TF-IDF.

\noindent \techname{Compute Similarity:}
 quantifying the similarity or proximity between events, sequences, or patterns via distance metrics.
 For example, Sequence Synopsis \cite{chen2017sequencesynopsis} uses Jaccard Index to calculate event similarity, and Levenshtein distance to calculate pattern similarity. 

 % \leo{we mentioned computing event similarity and computing embedding in Expand Attribute Set too. To me, Expand Attribute set only talks about how the derived information is stored, while the other strategies in this intent is about what is computed. I think we should focus on what, not how, and remove Expand Attribute Set. We can discuss under "Derive" how these new information is usually stored as new attributes (maybe in some other form too?)} \zinat{I merged the examples in expand attribute set to generate embeddings/ projections, and call it  obtain embeddings/ projections/ attribute and remove the discussion about storage}

\noindent \techname{Align Sequences:}
establishing correspondences between events across multiple sequences based on a common reference point or shared semantic meaning. Event sequence alignment is crucial when dealing with variable-length sequences, inconsistent event orders, or gaps. Event sequence tools typically align sequences using  a common origin time, such as first event. Some tools use specialized alignment algorithms, such as EventThread2 \cite{guo2018eventthread2} uses Dynamic Time Warping,
% , a technique that finds an optimal alignment between two sequences with variable lengths and event orders by allowing flexible mapping of events.
Sequence Braiding \cite{di2020sequencebraiding} finds the Shortest Common Supersequence.
% (SCS)  that contains all the events from the input sequences as subsequences. It ensures that the aligned sequences have a consistent event matching. 
% MatrixWave \cite{zhao2015matrixwave} inserts proxy events to serve as anchors for alignment. 

\subsubsection{ORGANIZE}
ORGANIZE refers to assembling and grouping similar sequences or events based on specific criteria or attributes.
% It involves grouping or clustering sequences that share common characteristics, patterns, or properties, enabling users to identify and analyze cohesive subsets within the overall dataset. The primary goal of organization is to provide a well-formed representation of the event sequences, facilitating the identification of similarities, differences, and relationships among groups of sequences.

\noindent \techname{Group Sequences:} organizing event sequences into distinct groups or clusters.
% based on shared attributes, properties, or derived characteristics.
\textit{Criteria} for grouping include:

\begin{itemize}[leftmargin=0pt]
\itemsep0em 
    \item[] \textbf{events}, e.g., group sequences by entry event (CoreFlow \cite{liu2017coreflow}, EventFlow \cite{monroe2013eventflow} ); event subsequences (milestones) or focal events of interest (GapFlow \cite{gotz2013gapflow})
    % use  to group sequences.
    \item[] \textbf{event properties}, to support attribute-based analysis, e.g., % Grouping by event occurrence time, occurrence intervals, duration, 
    group by categorical event types (MatrixWave \cite{zhao2015matrixwave}) or partitioning continuous event attributes into meaningful ranges (Segmentifier \cite{dextras2019segmentifier}).
    \item[] \textbf{sequence-level metadata}, such as demographics or other non-temporal attributes to support analyzing population subgroups. Common methods include grouping by sequence attributes: both categorical  like gender and numerical like age (TipoVis \cite{tiphan2015tipovis}); or derived sequence properties like total duration, event count (SessionViewer \cite{lam2007sessionviewer}).
\item[] \textbf{clustering on derived attributes}, involves  attribute-based grouping derived as discussed in  \ref{derive}.
Sequence Synopsis \cite{chen2017sequencesynopsis} clusters sequences mapping to the same pattern. EventThread \cite{guo2017eventthread}, groups sequence segments by cluster similarity. 
% Protosteer \cite{ming2019protosteer} groups sequences based on their proximity to learned prototypes.

% Protosteer \cite{ming2019protosteer} utilizes an interpretable sequence learning model, ProSeNet \cite{Ming_2019}, to learn and refine prototypes that capture the essential characteristics of the dataset. Sequences are then grouped together based on their similarity or proximity to these learned prototypes.
\end{itemize}

\noindent \techname{Create Event Hierarchy:} 
% \leo{Move this to Organize as we have discussed} \zinat{Moved} 
combining available event types into higher-level categories or meta-events based on semantic similarity, hierarchical relationships, or user-defined criteria. 
% The goal of event grouping is reducing the complexity and granularity of the data while still preserving the essential information. Grouping events into categories helps in managing large-scale event sequence data, facilitates analysis at different levels of abstraction, improves customization and flexibility.
The definition of event categories can vary based on criteria, such as: automated algorithms (e.g., Cadence \cite{gotz2019cadence}), existing hierarchical structures (e.g., MatrixWave \cite{zhao2015matrixwave}, Segmentifier \cite{dextras2019segmentifier}), regular expression (e.g., Eventpad \cite{cappers2018eventpad}) or user-defined criteria (e.g., Patterns and Sequences \cite{liu2016patternsandsequences}, Eloquence \cite{vrotsou2018eloquence}).

% Event sequence tools provide various features on their interfaces for users to create, modify, and manage event hierarchies. This may involve allowing users to introduce new event types at different levels of the hierarchy, assign existing event types to categories by dragging and dropping etc.

\subsubsection{SIMULATE}

Simulate refers to generating hypothetical or predictive scenarios, enabling users to analyze the potential outcomes, dependencies, and implications of different events or patterns. 
% The primary goal of simulation is to provide users with a means to experiment with the event sequences, test assumptions, and gain insights into the possible future states or alternative paths.

\noindent \techname{Mutate Components:}
altering specific components or properties to simulate and analyze alternative scenarios,
% providing insights into the sensitivity and robustness of subsequent events to changes to that specific event properties. This technique 
enabling what-if analysis.
% and exploring the impact of changing event attributes on the sequence outcome or downstream events. 
% The process involves applying modification to the selected event(s) and propagating the impact of the modified event to the subsequent events in the sequence,
In OpinionFlow \cite{wu2014opinionflow}, the impact of changing event attributes can be observed on downstream events via 
 adjusting tweet sentiment.
 % can help predict the potential influence on subsequent tweets.

% modifying specific components or properties to simulate and analyze alternative scenarios, enabling what-if analysis. In OpinionFlow [67], for example, sentiment analysis allows observing the effects of altering event attributes on downstream events, such as adjusting tweet sentiment
% \paragraph{Modify/Mutate Patterns and Analyze Effect}
% Modifying or mutating patterns refers to creating variations of existing representative patterns, and assessing the impact of pattern modifications on the overall analysis (ProtoSteer \cite{ming2019protosteer}).

\noindent \techname{Generate 
% Predict/Recommend 
Next Possible Events:} 
% \zinat{This is very similar to high-level goal prediction and recommendatiion}
predicting the likelihood of future events based on patterns observed in historical event sequences through  machine learning models. 
% The process includes training predictive models using the historical event sequences, and applying the trained models to generating predictions or estimates of the next likely events. 
Generating next possible events enables analysts to anticipate future outcomes, plan for potential scenarios, and make data-driven decisions
% based on the expected trajectory of event sequences 
(OpinionFlow \cite{wu2014opinionflow}).

\subsection{Simplify Data}
Data Simplification refers to reducing the complexity or scale of an event sequence dataset to make it more amenable to automated processing or human inspection.

\subsubsection{AGGREGATE}

Aggregate involves replacing multiple events with a single representative event, reducing the number or type of events in the visualization. 
% Creating event hierarchy to aggregate events into some super types reduces the number of distinct event types. Coalescing repeating events reduces number of events for long sequences. Aggregating events across sequences to create a tree or DAG structure reduces number of events in the entire dataset and helps in visualization when the number of sequences is high. It reduces the complexity and granularity of event sequences by combining related or repetitive events, thereby providing a more concise and manageable representation of the data. 

\noindent \techname{Coalesce Repeating Events into One:}
merging consecutive events  within a sequence into a single representative event.  
% Coalescing events helps in decluttering the visualization.
% by eliminating the representation of repeated events, improving readability, increasing scalability and comprehension of the sequences. 
The merging can be done by selecting one event as representative and discarding duplicates (e.g., Segmentifier \cite{dextras2019segmentifier}), or by creating a new event that encompasses the duration or attributes of the merged events (e.g., EventFlow \cite{monroe2013eventflow}). \looseness -1
% Coalescing is performed within a single sequence or pattern, and is helpful for condensing the number of events in long sequences with repetition events. 

% Event sequence tools may offer users the option to toggle the coalescing functionality on or off, allowing them to switch between the original and coalesced views (Segmentifier) . Coalescing can also be done on interval events, aggregating the time duration of the interval events of same type. (EventFlow).

% By identifying repeating events, merging them and updating the sequence representation, event coalescing simplifies the visual representation, reduces clutter, emphasizes essential patterns, and improves the scalability and interpretability of event sequence visualizations. 

\noindent \techname{Combine Events across Sequences:} aggregating similar events across multiple sequences and organizing them into a hierarchical or graph-based structure, such as a tree (e.g., EvenFlow \cite{monroe2013eventflow})  or a directed acyclic graph (DAG) (e.g., Sankey diagram). This transformation provides a more compact representation of the event sequences.
% (unsolved)\zinat{Do we need more details?} \leo{a figure will be helpful.}
% allowing for efficient exploration,and analysis of the data. By transforming the raw event sequences into a structured format, visualization tools can reveal patterns, frequencies, and relationships among events across different sequences.
% Aggregating events across sequences and representing them in a compact structure allows visualization tools to handle larger and more complex event sequence datasets with numerous sequences, as the transformed structure reduces the overall size of the data.

\subsubsection{SUMMARIZE} 

Summarize involves generating a concise overview of the entire dataset or its selected subsets. The process includes computing statistical distributions to present the data in an easily understandable form or applying algorithms to extract representative patterns. The primary goal of summarization is to provide a high-level understanding of the key characteristics and statistical properties without being overwhelmed by the full complexity of the data.

% Summarization helps with scalability and insight discovery.

\noindent \techname{Calculate Distribution:} apply statistical methods to form mathematical summaries of event sequences, enabling both confirmatory and exploratory analysis. These summaries encompass metrics on distributions and probabilities over event sequence attributes. Examples include histograms (e.g., Segmentifier \cite{dextras2019segmentifier}, Cadence \cite{gotz2019cadence}), type distributions (e.g., SessionViewer \cite{lam2007sessionviewer}), and inter-arrival times between event types (e.g., EventFlow \cite{monroe2013eventflow})

\noindent \techname{Extract Common Patterns:} \label{commonpatterns}
algorithmically extracting frequently occurring subsequences based on events that are present in the data. Examples include temporal pattern mining techniques, such as recursive branching pattern mining (e.g., CoreFlow \cite{liu2017coreflow}), VMSP (e.g., Patterns and Sequences \cite{liu2016patternsandsequences}, MAQUI \cite{law2019maqui}), Frequent Sequential Pattern (e.g., SentenTree \cite{hu2016sententree}), PrefixSpan (e.g., Eloquence \cite{vrotsou2018eloquence}), or Multiple Sequence Alignment (e.g., EventPad \cite{cappers2018eventpad}).
% Useful parameters include minimum support threshold, gap tolerances,etc..

\noindent \techname{Extract Latent Patterns:}\label{latentpatterns}
using statistical or machine learning methods to discover hidden sequential structures and trends implicitly embedded within event sequences. Examples include probabilistic methods, such as Hidden Markov Models (e.g., DPVis \cite{kwon2020dpvis}), Algorithms like hierarchical Bayesian Rose Tree models (e.g, OpinionFlow \cite{wu2014opinionflow})
% ,   CP decomposition (e.g, EventThread \cite{guo2017eventthread}) 
to group sequences with similar semantic content.
% Useful parameters include number of hidden states, threads,etc..

\subsubsection{SEGMENT}

Segment refers to breaking down lengthy sequences into smaller, more manageable segments based on specific criteria or temporal boundaries. 

\noindent \techname{Split Sequences or Patterns:}
% \leo{so there is only one technique for SEGMENT?} \zinat{Yes} 
decomposing event sequences into meaningful subsequences or segments.
% based on criteria, such as fixed time intervals, the occurrence of key events or patterns, or derived attributes. 
\textit{Criteria} for splitting include:

\begin{itemize}[leftmargin=0pt]
\itemsep0em 

\item[] \textbf{temporal folding}, dividing large-scale event sequence data collected over extended periods into segments of fixed time intervals, such as a year, a month, or a day) (e.g., EventThread \cite{guo2017eventthread}, STBins \cite{qi2020stbins}). 
% The choice of segmentation criteria depends on domain-specific understanding.

% Segment by Key Events: decomposing a sequence based on the occurrence of landmark events, patterns, or milestones that are considered significant for the analysis.
% This technique supports targeted pattern discovery, comparative analysis, and understanding of the role of key events in shaping trajectories. 
\item[] \textbf{key event}, dividing sequences into pre- and post-event segments to analyze antecedent and sequelae patterns relative to the key event (e.g., MAQUI\cite{law2019maqui}, EventThread \cite{guo2017eventthread}), or segmenting sequences based on the presence or absence of a key event or pattern to compare alternative trajectories ( e.g., DecisionFlow \cite{gotz2014decisionflow}, MAQUI \cite{law2019maqui})
\item[] \textbf{derived attributes}, such as content vector segmentation (e.g., EventThread2 \cite{guo2018eventthread2}).

\item[] \textbf{dynamic splitting} of segments with a shared pattern ( e.g., CoreFlow \cite{liu2017coreflow}, Patterns \& Sequences \cite{liu2016patternsandsequences}). 
\end{itemize}

\subsection{Configure Data} 
Configuring Data comprise modifying and experimenting with different components, properties, and granularity of an event sequence dataset for exploration, aiming to identify the optimal amount of data required for subsequent analysis. 
% The goal of data configuration is to provide users with the flexibility to adapt the data to their specific analysis need
% experiment with different data components and properties: refine, include/exclude, abstract/elaborate

\subsubsection{REFINE}

Refine refers to iteratively improving or fine-tuning the analysis data based on domain knowledge, or evolving requirements. 
% It involves making adjustments, modifications, or calibrations to various aspects of the analytic pipeline to enhance the accuracy, relevance, and usefulness of the event sequence analysis. 
Refinement enables progressive adjustments of analysis outcomes, incorporating user expertise and insights to obtain precise and meaningful results.

\noindent \techname{Modify Summary:}
supporting modifications and calibrations of  system-generated summarized representations or patterns, enabling users to incorporate their domain knowledge. 
% The event sequence visualization tools apply algorithms or techniques to generate initial summaries, and users examine the generated summaries to assess their relevance with their domain knowledge.
 Users can iteratively fine-tune the automatically generated summaries until they are satisfied with the representation.
 Examples of modification operations include adding/deleting/editing events in patterns (e.g., ProtoSteer \cite{ming2019protosteer}), and merging/splitting patterns (e.g., EventThread2 \cite{guo2018eventthread2}). 
Summaries produced by extracting common patterns (\ref{commonpatterns}) can be calibrated using minimum support thresholds and gap tolerances. Similarly, summaries generated from extracting latent patterns (\ref{latentpatterns}) can be customized by controlling  the number of hidden states.

% Summary Modification offers several benefits such as domain knowledge integration, increased interpretability.

\noindent \techname{Adjust Parameters:}
fine-tuning parameters that affect data processing and modeling,  without directly manipulating the visual representation itself.
Users can adjust various parameters, including temporal settings (e.g. window size or duration  in STBins \cite{qi2020stbins}), grouping settings (e.g., similarity measure thresholds  in EventThread \cite{guo2017eventthread}), aggregation settings (e.g., importance measures for grouping events in Cadence \cite{gotz2019cadence}), constraint/filtering parameters (e.g., toggling rules in EventPad \cite{cappers2018eventpad} or modifying constraints in Eloquence \cite{vrotsou2018eloquence}).

% or time granularity for temporal aggregation to control the level of abstraction (OpinionFlow \cite{wu2014opinionflow}). 
% The features for adjustment can include separate control panels, sliders, input fields, checkboxes, or dropdown menus, depending on the nature of the parameter. 
% Users can modify various parameters including workspace settings (e.g., toggling comparison mode in ProtoSteer [32]), temporal settings such as window size or duration (e.g., in STBins [40]), organization settings like similarity measure thresholds (e.g., in EventThread [16] or Maqui [26]), importance measures for grouping events (e.g., in Cadence [13]), and constraint/filtering parameters such as toggling rules or modifying constraints (e.g., in EventPad [5] or Eloquence [45]).

\subsubsection{INCLUDE-EXCLUDE}
Include-Exclude  refers to controlling the subset of event sequence data under consideration, enabling users to focus on relevant subsets and eliminate irrelevant information during analysis.
% It involves applying dynamic queries or cross-filtering techniques to selectively include or exclude specific events, sequences, or data points based on user-defined criteria. 
% The primary goal of include-exclude is to enable users to focus on relevant subsets of the data, eliminating irrelevant information at a specific point of the analysis.

% Do we need to separate dynamic queries and cross filtering
\noindent \techname{Execute Dynamic Queries:}
refer to applying filters, selections, or search criteria to retrieve matched subset.
% This technique allows analysts to explore and refine event sequences dynamically based on specific attributes, patterns, or user-defined constraints. This can be achieved on the system by providing interactive controls, such as checkboxes, sliders, search boxes, or lasso selection tools, that allow analysts to specify filtering criteria, and dynamically updating the visualizations to display only the event sequences or entities that meet the specified criteria, filtering out the irrelevant data points.
Dynamic queries span various components of event sequence data and visual elements, including filtering by event types, categories, or attributes (e.g., Cadence \cite{gotz2019cadence}, EventPad \cite{cappers2018eventpad}), selecting specific subsequences, stages, or time intervals (e.g., DPVis \cite{kwon2020dpvis}, Segmentifier \cite{dextras2019segmentifier}), searching for keywords or patterns (e.g., ProtoSteer \cite{ming2019protosteer}, Eloquence \cite{vrotsou2018eloquence}), filtering based on sequence attributes (e.g., Sequence Braiding \cite{di2020sequencebraiding}, Eloquence \cite{gotz2019cadence}), filtering based on transitions (e.g., DPVis \cite{kwon2020dpvis}), querying for the presence or absence of specific events, milestones, or patterns (e.g., Eloquence \cite{vrotsou2018eloquence}), and applying advanced querying techniques such as graphical queries, or regular expressions (e.g., Cadence \cite{gotz2019cadence}, EventFlow \cite{monroe2013eventflow}).

% Dynamic queries can be applied to various aspects of event sequence data and visual elements, such as:

% - Filtering by event types, categories, or attributes (Cadence, EventPad)
% - Selecting specific subsequences, stages, or time intervals (DPVis, ProtoSteer, Segmentifier)
% - Searching for keywords or patterns (ProtoSteer, Eloquence)
% - Filtering based on sequence attributes  (Sequence Braiding, Cadence, Eloquence)
% - Filtering based on transitions (DPVis)
% - Querying for the presence or absence of specific events, milestones or patterns (Eloquence)
% - Applying advanced querying techniques, such as graphical queries or milestone-based constraints, regular expressions (Cadence, Eloquence, )

% Dynamic queries are particularly useful in domains where event sequence datasets contain a large number of sequences. They allow analysts to focus on subsets of interest.

\noindent \techname{Cross-filter Components:} applying filters in one view and automatically updating other coordinated views.  By applying filters or selections in one context and observing the immediate impact on related contexts, analysts can gain insights into how different data attributes or dimensions correlate.
% and can be achieved by displaying multiple coordinated views, such as range charts for different variables  or synchronous update of data dimensions of the same underlying dataset, such as fetching only matched indices on the entire dataset based on condition on a specific dimension.
% \zinat{The examples are a bit odd, as they are talking about low level interaction techniques here, might need a rewrite }
Examples include cross-filtering via interactions (OpinionFlow \cite{wu2014opinionflow}, MAQUI \cite{law2019maqui}) or query languages (DPVis \cite{kwon2020dpvis}). 
% The system retrieves the subset of data based on selection, and it is propagated to all other coordinated views or data dimensions in real-time.

% Cross-filtering is particularly useful in domains where data has multiple dimensions or attributes, such as event sequences with metadata. It allows analysts to build cohorts, compare subgroups, and investigate the interplay between different data aspects.

\subsubsection{ABSTRACT-ELABORATE}
Abstract-Elaborate involves interactively adjusting the level of detail, granularity, or abstraction of event sequence data representation. This strategy enables dynamic exploration of different levels of abstraction, from high-level overviews to detailed, fine-grained representations. 
% The primary goal of abstraction and elaboration is to enable users to explore and understand the event sequence data at varying levels of detail, supporting both high-level pattern recognition and in-depth analysis of specific events or sequences.

\noindent \techname{Details-on-Demand:}
 providing users with additional information about specific event sequence components upon interaction. This technique helps analyst gain insights into specific aspects while maintaining the overall context. 
% \zinat{Same issue as cross-filtering, should we delete this line and keep the next? } 
% The process is supported by user interactions (e.g., DPVis \cite{kwon2020dpvis}, ProtoSteer \cite{ming2019protosteer}, MAQUI \cite{law2019maqui}, MatrixWave \cite{zhao2015matrixwave}, Cadence \cite{gotz2019cadence}).
This information can be presented in various forms, such as: tooltips (e.g., Segmentifier \cite{dextras2019segmentifier}), sidebar (e.g., STBins \cite{qi2020stbins}, OutFlow \cite{Wongsuphasawat2011Outflow}), separate information panels (e.g., Sequence Braiding \cite{di2020sequencebraiding}, EventPad \cite{cappers2018eventpad}) or Expanded views (e.g., DPVis \cite{kwon2020dpvis}).

% User interaction plays a crucial role in the process, supported by actions like hovering over (DPVis, ProtoSteer, Maqui, MatrixWave), clicking on, or selecting specific visual elements of interest (Cadence) such as events, sequences, patterns, nodes, or edges. This information can be conveyed through various formats including tooltips (Segmentifier), sidebars (STBins, Segmentifier, Maqui, OutFlow), separate information panels (Sequence Braiding, EventPad), or expanded views (DPVis).

\noindent \techname{Drill down-Roll up:} supporting  data navigation at various levels of detail. 
transitioning between detailed, lower-level representations (drill down) and summarized, higher-level abstractions (roll up). 
% This allows users to explore the event sequences at varying levels of abstraction, based on their analysis needs. 
For example, Lifeline \cite{plaisant1996lifelines} provides semantic zooming to expand desired facets (low-level abstraction), and use silhouettes and shadows when full details can't be shown on the overview  (high-level abstraction).
The technique is facilitated by underlying hierarchical data organization.  
% The visualization adapts to reflect the current level of detail based on the drill-down or roll-up operations. 
Implementation ideas include showing or hiding labels or annotations (e.g., Eloquence \cite{vrotsou2018eloquence}, OpinionFlow\cite{wu2014opinionflow}), creating a new context view (e.g., EventPad \cite{cappers2018eventpad}, Patterns and Sequences \cite{liu2016patternsandsequences}) or modifying the layout  (e.g., LifeLines \cite{plaisant1996lifelines}) to accommodate new level of detail.

\subsection{Configure Visualization}
Configuring visualization refers to modifying and adjusting different visual properties, and layouts of an event sequence visualization to explore alternative representations, and enhance the communication of insights. Visual configuration allows users to tailor the visualization to their perceptual preferences and storytelling needs.
% by iteratively refining the visual representation to facilitate knowledge dissemination.

\subsubsection{VISUALIZE}
Visualize refers to applying or updating rules that map data to visual marks or mark properties, creating graphical representations. 
% It involves transforming event sequence data into visual encodings that effectively communicate patterns, relationships, and insights to the users.

\noindent \techname{Produce Visual Representation:}
converting sequences and associated attributes into visual forms that effectively communicate patterns, relationships, and insights. The generated visual representations leverage various visual elements, such as timelines, graphs, charts, glyphs, and diverse color schemes, to convey the structure, temporal dynamics, and characteristics of the event sequences.

The data elements are mapped to visual channels such as position, size, color, shape, or opacity. The choice of visual encodings depends on the nature of the data, the analysis goals, and the desired level of detail or abstraction. A detailed description of all kinds of visual representation in beyond the scope of this work, and has been covered extensively in prior works \cite{jentner2019techniquesforpatterns, zinat2023visual}.

\noindent \techname{Customize Visual Encoding:}
directly modifying visual representations of event sequences and associated attributes. The primary focus is enhancing readability, interpretability, and communication of insights by aligning visual representations with user preferences or analysis goals. \looseness - 1
% It involves adapting the visual variables, such as color schemes, node sizes, edge thicknesses, or glyph designs, to effectively communicate the desired information and insights. 
% The primary focus is to enhance readability, interpretability, and communication of insights by aligning the visual representation with the user's preferences or analysis goals.
% Modifying visual encodings directly impacts how the event sequences are visually presented to the user.

Customization may include visual encoding adjustments such as line thickness or spacing (e.g., sequence braiding \cite{di2020sequencebraiding}), selecting different visual designs (e.g., ActiviTree \cite{vrotsou2009activitree}, STBins \cite{qi2020stbins}), adjusting transparency levels (e.g., Eloquence \cite{vrotsou2018eloquence}), changing canvas size (e.g., CoreFlow \cite{liu2017coreflow}), selecting color palettes (e.g., Eloquence \cite{vrotsou2018eloquence}), defining node shapes, edge styles, or glyph designs.
% that best represent the event types, attributes, or relationships within the sequences.  Users can also adjust node sizes to reflect event durations (Eloquence) or frequencies, or vary line  thicknesses (Sequence Braiding) 

% Customizing visual encodings allows users to optimize the readability and interpretability of the visualizations that align with specific analysis goals. By adjusting parameters such as spacing, scale, or visual clutter reduction, users can create visualizations that are easier to perceive, navigate, and understand, even when dealing with complex or large-scale event sequence data.

\noindent \techname{Update Layout:} adapting the presentation mode of an event sequence visualization to accommodate diverse analysis requirements. It involves adjusting layout parameters, such as canvas dimensions (e.g., CoreFlow \cite{liu2017coreflow}) or visualization modes (e.g., $($s|q$)$ueries \cite{zgraggen2015squeries}, ProtoSteer \cite{ming2019protosteer}).

\subsubsection{NAVIGATE}
Navigate depicts interactively exploring and modifying viewport or visible area of a visualization to focus on specific regions or time periods. \looseness - 1 
% The goal of navigation is to provide users with the ability to dynamically adjust their view of the data, enabling them to gain both holistic overviews as well as fine-grained insights.

\noindent \techname{Zoom:}
dynamically adjusting the  magnification level within a single, consistent view.
% It involves adjusting the magnification or resolution of the displayed information. 
Zooming enables multi-scale exploration to ensure readability and clarity of the displayed information.
Visualization tools may employ  animated transitions or progressive loading to enhance the zooming experience.
% Users can zoom in to see more details, such as enlarging the size of events, labels, or visual markers, or zoom out to gain a broader overview of the entire event sequence dataset. Zooming helps users to examine specific regions, patterns, or time ranges without fundamentally changing the granularity or hierarchical level of the data. 
Users can initiate the zooming action through 
% various interaction techniques, such as using
dedicated zoom controls (e.g., MatrixWave \cite{zhao2015matrixwave}, LifeLines2 \cite{wang2008aligning})
% (LifeLines \cite{plaisant1996lifelines}, EventThread \cite{guo2017eventthread}, LifeLines2 \cite{wang2008aligning}, LifeFlow \cite{wongsuphasawat2011lifeflow}, EventFlow \cite{monroe2013eventflow}, MatrixWave \cite{zhao2015matrixwave}, Sequence Braiding \cite{di2020sequencebraiding})
% ,  or selecting a specific region or range of interest/ (e.g., EventThread2 \cite{guo2018eventthread2}).

\noindent \techname{Pan:} interactively exploring different parts of the displayed event sequences by shifting the viewport.
% It involves providing mechanisms for users to smoothly scroll, slide, or move the view to reveal additional information currently outside the visible bounds of the display. 
Panning enables inspection of lengthy or complex sequences in a continuous and fluid manner without losing context.
% Panning is supported through various interaction techniques, such as clicking and dragging. 
Usually, horizontal panning shifts the visible area along the time axis, revealing the temporal progression and relationships among events (e.g., LifeLines2 \cite{wang2008aligning}, MatrixWave \cite{zhao2015matrixwave}). Some tools also support vertical panning, especially when dealing with a large number of sequences (e.g., EventPad \cite{cappers2018eventpad2}). 

 % Panning allows users to explore event sequences that extend beyond the initial visible area of the display. By providing smooth and responsive panning capabilities, the visualization tool empowers users to examine lengthy sequences and seamlessly navigate through the data. Panning facilitates data-driven exploration, enabling users to uncover insights and relationships that may not be immediately apparent in a static view.

\subsubsection{FOCUS} 
Focus denotes visually emphasizing specific elements, patterns, or subsets of interest, guiding user attention towards the emphasized element. \looseness -1

\noindent \techname{Highlight Marks:}
emphasizing or accentuating specific components of interest within the visualization. This involves applying distinct visual attributes or effects to the selected marks, such as changing color (e.g., SentenTree \cite{hu2016sententree}), opacity (e.g., MatrixWave \cite{zhao2015matrixwave}, SentenTree \cite{hu2016sententree}), adding border (e.g., Lifelines \cite{plaisant1996lifelines}, Patterns and Sequences \cite{liu2016patternsandsequences}), or using visual cues like glyphs (e.g., TipoVis \cite{tiphan2015tipovis}) or annotations.
% Highlighting marks helps users focus their attention on important aspects of the event sequences, identify relationships or dependencies, and facilitate comparative analysis.
% Selection: Users interact with the visualization to select the marks they want to highlight through various mechanisms, such as clicking on individual events, brushing a range of events, or specifying query criteria. Visual emphasis: Once the marks are selected, the tool applies visual emphasis to make them stand out from the rest of the visualization. 
% By visually emphasizing specific marks, the tool helps users to identify and locate important information within the potentially large and complex event sequence data.

\noindent \techname{Select \& Link Components:}
synchronized selection and highlighting of elements across different views.
% In other words, user selection on one view, automatically selects or highlights corresponding elements in other views. 
By visually connecting related elements across views, linked selection helps users gain a holistic understanding.
% This task enables users to develop a coherent understanding of the relationships and dependencies between different aspects of the data.
% Users interact with an element in one view of the visualization, such as clicking on a pattern in a summary view, hovering over a cluster node in a cluster view, or selecting a user from a user list view. The tool identifies the elements in other views that are related to the selected or hovered element based on data attributes.  The corresponding or related elements in the other views are automatically highlighted or selected to visually indicate their connection to the initially selected element. The highlighting can be done through various visual cues, such as changing color, opacity, or adding visual markers or annotations. The linked selection/highlighting is coordinated across multiple views or representations of the event sequences. 
For example, in Sequence Synopsis \cite{chen2017sequencesynopsis} selecting a pattern in the summary view highlights individual sequences containing the pattern in the detailed view. In DecisionFlow \cite{gotz2014decisionflow}, selection of an event in other panels remain consistent across pattern panel. 

% Linked selection/highlighting provides users the context of the selected element to other aspects of the data. By visually connecting related elements across views, linked selection/highlighting helps users derive insights. and gain a holistic understanding. It enhances the usability of the visualization tool by reducing the need for manual navigation and search.

\subsubsection{REARRANGE}

Rearrange depicts changing the spatial arrangement or positioning of visual items within the visualization layout. It involves modifying the order  of visual elements to improve overall readability and interpretability. \looseness -1
% of the event sequence visualization.
% change layout
% Merged with Adjust Parameters

\noindent \techname{Reposition Sequences:}
 changing the
 % baseline of the displayed sequences or patterns based on a selected event. This technique adjusts the 
 visual arrangement by centering sequence layout around a specific event of interest, facilitating exploration of temporal relationships before and after the alignment point. Repositioning can help uncover insights that may be obscured in default configuration. This technique is often accompanied by an animated transition for smooth user experience (e.g., Sequence Synopsis \cite{chen2017sequencesynopsis}).
% The alignment event can be selected by clicking on a specific event in a summary view (e.g., Eloquence \cite{vrotsou2018eloquence}, Patterns and Sequences \cite{liu2016patternsandsequences}), choosing an event type from a dropdown menu (e.g., LifeFlow \cite{wongsuphasawat2011lifeflow}, EventFlow \cite{monroe2013eventflow}), or specifying a custom alignment condition. 
% (unsolved)\leo{we may want a figure to clearly show the diff between align sequences and reposition sequences}
% The sequences are then visually rearranged so that the selected event becomes the central reference point, and all other events are positioned relative to it. 
 % By allowing users to reposition the sequences based on specific events, the tool enables them to focus their analysis on the events that are most relevant or interesting to their exploration. Users can align the sequences based on critical events, decision points, or other significant milestones in the data.

\noindent \techname{Reorder/Sort Components:}
reordering the displayed components, in ascending or descending order.
% By allowing users to sort based on specific criteria, the tool enables them to prioritize and focus on the most relevant components of the data 
% The sorting criteria can be chosen interactively through a dropdown menu (Lifelines2), a set of radio buttons (Lifelines2), drag and drop (SessionViewer, EventPad) or other interactive controls that provide sorting options. The sequences or events are visually reordered vertically so that they are displayed in ascending or descending order according to the selected metric or attribute. Users can interactively change the sorting criteria or switch between different sorting options to explore the data from multiple perspectives. 
Sorting across multiple attributes is possible, with primary criteria taking precedence and subsequent ones breaking ties or refining the ordering.
Sorting can be performed on subsets, such as sorting sequences within a cluster, events within a time range, or attributes within a category.
% Sort is an essential task in event sequence visualization that allows users to reorder the displayed sequences, events, or attributes based on specific criteria or metrics.

There is a wide range of sorting criteria, including, frequency (LifeFlow \cite{wongsuphasawat2011lifeflow}), number of occurrences (TipoVis \cite{tiphan2015tipovis}), rarity (EventPad \cite{cappers2018eventpad}),
% number of sequences in the corresponding cluster(EventPad \cite{cappers2018eventpad});
similarity (Sequence Synopsis \cite{chen2017sequencesynopsis}), alphabetical order (MatrixWave \cite{zhao2015matrixwave}), average time to previous event (LifeFlow \cite{wongsuphasawat2011lifeflow}), correlation with outcome (Cadence \cite{gotz2019cadence}), accuracy or performance (ProtoSteer \cite{ming2019protosteer}), attribute values (quantitative) (Sequence Braiding \cite{di2020sequencebraiding}), attribute name (qualitative) (TipoVis \cite{tiphan2015tipovis}), custom (MatrixWave \cite{zhao2015matrixwave}).

% (unsolved) \zinat{should we skip this long discussion?} \leo{condense it into a single paragraph.}

% Frequency-based: Number of sequences that include the event type (LifeFlow \cite{wongsuphasawat2011lifeflow}, Cadence \cite{gotz2019cadence}), total number of occurrences of an event type  (Lifelines 2 \cite{wang2008aligning}, TipoVis \cite{tiphan2015tipovis}, MatrixWave \cite{zhao2015matrixwave}), rarity of events (EventPad \cite{cappers2018eventpad}), number of sequences in the corresponding cluster(EventPad \cite{cappers2018eventpad})

% Alphabetical Criteria: Alphabetical order (Lifelines 2 \cite{wang2008aligning}, MatrixWave \cite{zhao2015matrixwave}, EventPad \cite{cappers2018eventpad})

% Temporal Criteria: Average time to previous event (LifeFlow \cite{wongsuphasawat2011lifeflow})

% Correlation-based Criteria: Correlation between the occurrence of an event type and the outcome label (Cadence \cite{gotz2019cadence})

% Prediction-based Criteria: Sorting based on the accuracy or performance of prediction results (ProtoSteer \cite{ming2019protosteer})

% Attribute-based Criteria: Attribute values (quantitative) (Sequence Braiding \cite{di2020sequencebraiding}), specific Event Attribute Name (qualitative) (SessionViewer \cite{lam2007sessionviewer}, TipoVis \cite{tiphan2015tipovis})

% Similarity-based Criteria: Similarity between patterns measured through editing distance (Sequence Synopsis \cite{chen2017sequencesynopsis})

% User defined Criteria: MatrixWave \cite{zhao2015matrixwave}, EventPad \cite{cappers2018eventpad}

\subsection{\rebuttal{Manage} Provenance}

Provenance encapsulates tracking, recording, and managing the history of data states, insights, and actions throughout exploration and analysis. It involves documenting applied transformations and their validations, result interpretations, and decision-making processes. The aim is to ensure transparency, reproducibility, and trustworthiness of analysis results by providing a comprehensive trail of the steps taken, assumptions made, and conclusions drawn. 

\subsubsection{DOCUMENT}

Documenting involves providing the means to create, manage, and organize insights or metadata associated with analysis.
% specific events sequence components within the visualization.
% facilitating the tracking, retrieval, and communication of relevant information. The primary goal of bookkeeping is to support users in documenting their findings, insights, or hypotheses, enhancing the interpretability and reproducibility of the event sequence analysis.

\noindent \techname{Annotate Components:}
providing mechanisms to add textual notes, comments, or descriptions to specific elements, subsets, or regions of the visualized data. This allows users to document their findings, insights, hypotheses, or any relevant information discovered during the analysis process. Annotation facilitates collaboration, knowledge sharing, enabling future reference (e.g., SessionViewer \cite{lam2007sessionviewer}).
% , a pop-up dialog box, or an inline editing functionality that allows users to type or edit the annotation directly within the visualization. The tool maintains a mapping between the annotations and the corresponding elements, subsets, or regions of the visualized data. The annotations can be visually displayed in proximity to the associated target or in a designated annotation panel within the user interface. It can be shown as text labels, tooltips, or expandable notes that provide additional details or context.

\noindent \techname{Save/ Record Analysis:}
storing specific snapshots of the analysis process for future reference or further investigation. Save/Record enables iterative analysis, and ensures reproducibility. This technique enables users to create a record of their analytical progress (e.g., Segmentifier \cite{dextras2019segmentifier}, ProtoSteer \cite{ming2019protosteer}), key findings, or interesting subsets of the data (e.g., EventPad \cite{cappers2018eventpad}, DPVis \cite{kwon2020dpvis}),as well as  design preferences (e.g., Eloquence \cite{vrotsou2018eloquence}) facilitating the ability to resume the analysis at a later time, or compare different stages of the exploration.
 
% This operation can be through a dedicated "Save" or "Record" button, a context menu option, or a keyboard shortcut.
% The ability to save, export, and share snapshots of the analysis also promotes collaboration and knowledge sharing among users. 

\noindent \techname{Insert New Marker Event:}
\rebuttal{manual insertion of} new events at specific points within existing event sequences. These user-defined events, known as marker events, serve as reference points  deemed significant by the users for analysis or interpretation (e.g., EventFlow \cite{monroe2013eventflow}, GapFlow \cite{gotz2013gapflow}). By creating new marker events, users can enrich the event sequences with additional contextual information.

% Marker event creation can be done by clicking on a particular position or specifying a timestamp and providing the necessary details and attributes for the new marker event, such as specifying a name or label for the event, assigning a category or type, adding descriptive annotations etc.

% \zinat{Data Preprocessing is a crucial part of event sequence analysis. Overview of the entire dataset is more important than details (Shneiderman Mantra)
% Data preprocessing helps with overview construction
% Automated mining techniques may have parameters not directly controllable in UI, but these mining parameters control the quality of summary.}

% \zinat{visualizations should explain relationships between patterns and sequences. Different representations may be needed for different pattern types and properties.
% 1. Goal
% \paragraph{many to many mapping with intent}
% 2. Intent
% 3. Technique}

% \zinat{1. Realized each paper has a overarching high level goal assicated with it, such as earlier tools like LifeFlow, Lifeline facilitater pattern exploration in general, matrixWave, Coco suppport cohort comarison,  DPVis models disease progression pathway. Thus We identified six such goals Cohort comparison
% Anomaly detection
% Identify common behavior
% Stage Progression
% Temporal Pattern Exploration 
% Sequence Modeling: Prediction, Correlation, Causality}
\section{Evaluation of the Framework}\label{sec:eval}

% \zinat{Link with Intro}
In this section, we evaluate  
% demonstrate the ability of our multi-level task framework to capture the wide range of tasks, intents, strategies, and techniques in event sequence analysis. We validate  our framework by comparing it with 
% in contrast to the limitations of 
existing task taxonomies through case studies. We then discuss future work to investigate the framework's evaluative and generative powers.  

\subsection{Descriptive Power} 
 \begin{table*}[!t]\fontfamily{ptm}\selectfont\small
\setlength{\belowcaptionskip}{-15pt}
\caption{Comparative analysis of task mappings for three case study excerpts from SeqCausal \cite{jin2021seqcausal}, RASIPAM \cite{wu2022rasipam} and FlexEvent \cite{linden2023flex} using our multi-level task framework and existing task taxonomies (Plaisant et al. \cite{plaisant2016diversity}, Du et al. \cite{du2016volumeandvariety}, Peiris et al. \cite{peiris2022data}). The color scheme indicates the level of alignment: \textcolor{Red3}{no match} and \textcolor{magenta}{partial match}. Our framework demonstrates more comprehensive mapping compared to existing taxonomies}
\label{tbl:case1}
% \centering
{\small
\begin{tabular}{p{0.025\linewidth}
p{0.135\linewidth}
p{0.14\linewidth}
p{0.15\linewidth}
p{0.135\linewidth}
p{0.14\linewidth}
p{0.12\linewidth}}
\hline
\textbf{Tasks} & \textbf{C1.T1} \cite{jin2021seqcausal} & \textbf{C1.T3} \cite{jin2021seqcausal} & \textbf{C1.T7} \cite{jin2021seqcausal} & \textbf{C2.T7} \cite{wu2022rasipam} & \textbf{C2.T9} \cite{wu2022rasipam} & \textbf{C3.T7} \cite{linden2023flex} \\
\hline
\textbf{Excerpt} & \textit{The doctors queried a group of 127 middle-aged patients aging from 50 to 60 who were diagnosed with pneumonia.} & \textit{After several iterations of confirming causalities and model updates,} $\dots$ & \textit{The doctors saved the final causality to the analysis history view.} & \textit{E2 re-ranked the tactics in Tactic View based on the tactical importance} $\dots$ & \textit{Experts applied this merging adjustment and obtained a more accurate estimate of the win rate for this serving tactic.} & \textit{Changing the color attribute to sepsis}, $\dots$ \\
\hline
\textbf{Plaisant et al. \cite{plaisant2016diversity}} & \textit{Prepare or select data for further
study}\newline
Identify a set of records of interest & 
\textit{Prepare or select data for further
study}\newline
Review data quality and inform choices to be made in order to model the data & \textcolor{Red3}{n/a} & \textcolor{magenta}{\textit{Prepare or select data for further
study}\newline Identify a set of records of interest} &
\textit{Prepare or select data for further
study}\newline
Review data quality and inform choices to be made in order to model the data &  \textcolor{Red3}{n/a} \\
\hline
\textbf{Du et al. \cite{du2016volumeandvariety}} & \textit{Extraction Strategies}\newline Goal-Driven Record Extracting & \textcolor{Red3}{n/a} & \textcolor{Red3}{n/a} & \textcolor{Red3}{n/a} & \textcolor{Red3}{n/a}  & \textcolor{Red3}{n/a} \\
\hline
\textbf{Peiris et al. \cite{peiris2022data}} & action: Filter\newline target: Event Sequences\newline criteria: Metadata Attributes & action: Derive Metrics\newline target: \textcolor{Red3}{n/a}\newline criteria: \textcolor{Red3}{n/a} & action: \textcolor{magenta}{Annotate}\newline target: \textcolor{Red3}{n/a}\newline criteria: \textcolor{Red3}{n/a} & action: {Sort/Rank}\newline target: \textcolor{magenta}{Event Sequences}\newline criteria: Metrics/Features & action: Add/Modify \textcolor{magenta}\newline target: \textcolor{magenta}{Event Sequences}\newline criteria: Metrics/Features & \textcolor{Red3}{n/a} \\
\hline
\textbf{Ours} & Intent: Configure Data\newline Strategy: Include-Exclude\newline Technique: Execute Dynamic Queries\newline action: Query\newline input: Event Sequences\newline output: Filtered Event Sequences\newline criteria: Age & Intent: Configure Data\newline Strategy: Refine\newline Technique: Adjust Parameters\newline action: Adjust\newline input: Current causal model\newline output: Updated causal model\newline criteria: Domain knowledge & Intent: Manage Provenance\newline Strategy: Document\newline Technique: Save/Record Analysis\newline action: Save/Record\newline input: Analysis State\newline output: Saved/Recorded Analysis\newline criteria: User-specified snapshot & Intent: Configure Visualization\newline Strategy: Rearrange\newline Technique: Reorder/Sort Components\newline action: Reorder/Sort\newline input: Tactics\newline output: Reordered Tactics\newline criteria: Tactical importance metric & Intent: Configure Data\newline Strategy: Refine\newline Technique: Modify Summary\newline action: Modify\newline input: Tactics\newline output: Modified Tactics\newline criteria: Domain knowledge  & Intent: Configure Visualization\newline Strategy: Visualize\newline Technique: Produce Visualization\newline
action: Produce \newline input: Event sequence data \newline
 output: Visual representation \newline
 criteria: Visual encoding rules\\
\hline
\end{tabular}}
\vspace{-1mm}

\end{table*}
 
 \rebuttal{To showcase the descriptive power of our multi-level task framework, we apply our framework and three existing frameworks \cite{peiris2022data,du2016volumeandvariety,plaisant2016diversity} to analyze real-world case studies reported in the literature. We use the following criteria to choose the case studies: 1) they should cover different application domains and analysis objectives, 2) are preferably published in recent years to reflect current challenges, and 3) are preferably not included in our coding and derivation of the framework. Based on these criteria, we chose the following three case studies: \textbf{C1}: causality in electronic health records from SeqCausal \cite{jin2021seqcausal}, \textbf{C2}: tennis tactics analysis from RASIPAM \cite{wu2022rasipam}, \textbf{C3}:  neonatal data similarity analysis from FlexEvent \cite{linden2023flex}.}

\rebuttal{\textbf{C1} and \textbf{C3} both were conducted in the healthcare domain, but they address different \texttt{objectives}: ``Correlation \& Causality'' (\textbf{C1}) vs. ``Pattern Exploration'' (\textbf{C3}). Conversely, \textbf{C2}  and \textbf{C3}, though addressing the same \texttt{objective} (``Pattern Exploration''), come from different domains: healthcare (\textbf{C3}) and sports analytics (\textbf{C2}). All the selected case studies were published in the last three years: \textbf{C1} (2021), \textbf{C2} (2022), and \textbf{C3} (2023). Among these three, only the paper for \textbf{C1} is included in our original corpus, while papers for \textbf{C2} and \textbf{C3} are not part of the original corpus. The total number of tasks is 25 across the case studies} \looseness -2

\rebuttal{For each case study, we coded the analysis process using our framework and three existing frameworks. Two authors independently performed the coding for each case study across all four frameworks, then compared results and reached a consensus through discussion. This thorough examination demonstrates our framework’s ability to comprehensively describe and characterize complex tasks and workflows. Here we present some key examples from the case studies. The full coding is added to the supplemental materials.}

\subsubsection{Takeaways from Three Case Studies}
 The comparative analysis of task mapping across the three case studies reveals several key differences among the approaches. We present six example mappings from the case studies in \cref{tbl:case1}.
 % \leo{if we include these examples, we can refer to the row number/task number in the content below.} Please refer to the Supplementary Materials for all the comparison results.

 \begin{description}[leftmargin=0pt]
\itemsep0em 
     \item[Granularity and Specificity:] Plaisant et al.'s  taxonomy \cite{plaisant2016diversity} provides only high-level descriptions of tasks, lacking low-level details that capture individual steps of analysis workflows. This limitation becomes evident when diverse low-level actions, such as grouping, sorting and filtering,  are required to accomplish the same high-level task, e.g., `Identify a set of records of interest ' (\textbf{C1.T1, C2.T7}). \\
     % In contrast, our framework offers a more fine-grained and precise characterization of tasks, enabling a deeper understanding of the analysis process.\\
     Similarly, Peiris et al.'s taxonomy \cite{peiris2022data} exhibits limitations in granularity for certain tasks, 
     % \leo{not sure what partial match or full match mean exactly. To me, this example is also about granularity. We can merge this with the first point?} 
     indicating the need for a more comprehensive and precise task framework. For example, both updating a model and performing dimensionality reduction tasks are mapped to action `Derive Metrics' (\textbf{C1.T3}). This mapping fails to capture the distinct nature of these tasks, as updating a model involves refining an existing model, while dimensionality reduction focuses on projecting data to a lower-dimensional space. In contrast, our framework captures this complexity through two levels of abstraction: `Refine' as strategy, and `Adjust Parameters' as technique for model update, and `Derive' as strategy and `Obtain Embeddings/Projections/Attributes' as technique for dimensionality reduction (\textbf{C1.T3}). These mappings align more closely with the actual tasks performed in the respective case studies.
     \item[Comprehensiveness:] Du et al.'s strategies \cite{du2016volumeandvariety} primarily addresses tasks related to reducing volume and variety, limiting their applicability to the full range of tasks involved in event sequence analysis. This narrow scope is apparent: tasks such as updating models, managing provenance, grouping sequences and refinement of the analysis are not  captured by their strategies (\textbf{C1.T3-C3.T7}).\\
     \rebuttal{In contrast, our framework captures tasks that are not adequately represented in the other taxonomies, showcasing its compatibility in covering the end-to-end tasks involved in event sequence analysis.} 
     % From initial data processing (e.g., obtain projections in \textbf{T6}), to data and visual configuration (e.g., modify summary in \textbf{T5}, change visual encoding in \textbf{T7}), to provenance (e.g., save analysis in \textbf{T3}). 

     However, It is important to acknowledge our technique list is not exhaustive, leaving room for  extension.  This is evident in \textbf{Task C3.T6 in the  supplemental materials (Table 3)}, where our framework identifies only a partial match. In this specific task, the grouping operation is being conducted on projection embeddings rather than event sequences. While our framework correctly identifies the action as `Group', it does not provide a perfect match for the input of the grouping operation.
     % Our framework, on the other hand, comprehensively covers a wide array of tasks, including those beyond data reduction, providing a more complete representation of the event sequence analysis workflow.
     % \item[Partial Matches:] Peiris et al.'s taxonomy \cite{peiris2022data} exhibits limitation in finding full matches for certain tasks or provides only partial matches, \leo{not sure what partial match or full match mean exactly. To me, this example is also about granularity. We can merge this with the first point?} indicating the need for a more comprehensive and precise task framework. For example, For example, tasks such as updating a model and performing dimensionality reduction are both mapped to the same action, ``Derive Metrics''. This mapping fails to capture the distinct nature of these tasks, as updating a model involves refining an existing model, while dimensionality reduction focuses on projecting data to a lower-dimensional space. Our framework captures this complexity through two levels of abstraction: derive as the strategy, and \leo{name the techniques} as the techniques.
     \item[Triplet vs. Quartet:]  \rebuttal{Peiris et al. \cite{peiris2022data} mention \textit{criteria} as a data component, limiting its ability to fully address the \textit{how} aspect of task characterization. In contrast, \textit{criteria} in our framework  may capture both the methods and conditions of an action, offering a more complete
characterization of how tasks are performed.
% \leo{metnion what we do} 
Additionally, their \textit{action-target-criteria} triplet lacks an \textit{output} component, obscuring important details about the results.
% the absence of \textit{output} component in their \textit{action-target-criteria} triplet obscures important details about the generated results. 
For instance, when analyzing summary patterns, it is crucial to discern whether these patterns are directly observable or represent latent structures requiring further interpretation.
% The omission of \textit{output} component makes it challenging to fully comprehend the implications of the generated results.
Our framework includes the output component, precisely distinguishing latent and common patterns.}
% In comparison, our framework incorporates \textit{output} as a component,  therefore precisely distinguishing latent and common patterns.
\rebuttal{Another notable difference is the handling of \textit{targets} and \textit{criteria}. Peiris et al. 
categorize four target types and five criteria types. 
% expresses \textit{targets} and \textit{criteria} as a fixed set of data characteristics, with four types of targets and five types of criteria. 
However, we believe that \textit{input} (analogous to \textit{targets}) and \textit{criteria} are too expansive to be limited to a predefined set of elements.} \looseness -2
% and captures the methods of an action through the criteria dimension, providing a more complete and detailed characterization of how tasks are performed. 
     % Our framework addresses these limitations by providing more accurate and specific mappings, such as "Adjust Parameters" for the model update task and "Modify Summary" for the merging adjustment task.
     % \item[Accuracy and Precision:] Our framework consistently offers more accurate and precise mappings for various tasks compared to the other taxonomies. For example, our framework maps saving analysis history to `Save/ Record', compared to `Annotate' by Peiris et al \cite{peiris2022data}. (\textbf{T3}) \leo{these examples have been used in previous points, consider using new examples?}
     % \item[Comprehensiveness:] \leo{how is this different from scope and coverage? considering merging them.}  Our framework captures tasks that are not adequately represented in the other taxonomies, showcasing its compatibility in covering the end-to-end tasks involved in event sequence analysis. From initial data processing (summarize, group), to data and visual configuration (modify summary, change visual encoding), to provenance (save analysis). 
     % Examples include the provenance task in case study 1 and the changing color attributes task in case study 3. 
     
 \end{description}

The comparative analysis of the task taxonomies across the three case studies highlights the strengths of our multi-level task framework.
% in terms of its granularity and comprehensiveness. 
By encompassing a broad spectrum of tasks, our framework enables a  holistic tracing of the event sequence analysis workflow. 
% By providing a more precise and inclusive characterization of tasks, our framework facilitates a deeper understanding of the event sequence analysis process and supports the comparison and generalization of insights across different domains. The case studies demonstrate the framework's ability to capture the complexities and nuances of real-world event sequence analysis workflows, validating its descriptive power and practical utility.

\subsection{Evaluative Power}

% While the focus of this paper is on the descriptive and evaluative aspects, we acknowledge the potential for generation power as a promising direction for future work. Our multi-level task framework lays the groundwork for the development of formal task specifications and technique recommendation systems. 
% By providing a structured and comprehensive representation of tasks and techniques in event sequence analysis, our framework can guide the creation of domain-specific languages, task-driven system design, and intelligent assistance for analysts working with event sequence data.

% By providing a comprehensive typology of objectives, intents, strategies, and techniques, 
Our framework lays the foundation for  structured evaluation of the effectiveness and completeness of event sequence analysis tools across diverse domains. Researchers can compare their capabilities by aligning the supported \objectives, \intents, \strategies, and \techniques \ of each tool to our framework levels, identifying commonalities, differences, and potential gaps. This comparative analysis can reveal strengths and limitations of individual tools and highlight areas for improvements.
% or additional features are needed. 
Moreover, the framework can provide a base for defining evaluation criteria and metrics to assess performance and usability.
% of event sequence analysis tools. 

% However,  it is important to note that
Further research and validation are necessary to fully establish the framework's evaluative capabilities. While the framework provides a comprehensive structure for  assessment, its effectiveness in practice needs to be demonstrated through case studies and applications in diverse domains, which are beyond the scope of this paper. 

\subsection{Generative Power}

Similar to Brehmer and Munzner's  typology \cite{brehmer2013typology}, our framework has the potential to guide practitioners through the `discover' and `design' stages of design studies \cite{sedlmair2012design}. 
% in a data-driven manner. 
% During the `discover' stage, 
Practitioners can translate their domain problems into abstract task descriptions by utilizing the \texttt{objectives}  outlined in our framework, therefore identifying high-level analysis goals that need to be supported.

% \zinat{is there any papers that confirm data analysis is a part of design study?}

% \zinat{What I want to convey in this paragraph, is first steps of the analysis are data-driven, and they can be mapped to our intents. On the other hand, final steps of the analysis are goal-driven, they can also be mapped to our intent. But the writing here requires a revision}

The hierarchical nature of our framework also allows practitioners to focus on high-level \intents \ without worrying about low-level details. For example, 
% Practitioners can then break down the analysis into granular steps by identifying the intents, connecting data to abstract tasks. The choice of intents can depend on the characteristics of the underlying dataset or the specific goals of the analysis. For example, 
if the dataset is large and complex, the initial analysis steps may require ``Data Simplification'' to make the data more manageable. Similarly, if the analysis goal is ``Prediction \& Recommendation'', practitioners should consider if data augmentation is necessary first.
% `Pattern Exploration', the final steps may also involve `Data Simplification' to extract meaningful patterns from the event sequences.

Once an intent has been identified, our framework offers an array of \strategies \  to support these \intents \  based on data characteristics. For instance, both `Include-Exclude' and `Abstract-Elaborate' strategies  cater to the `Configure Data' intent, yet they serve different purposes.
% `Abstract-Elaborate' supports focusing on more or less data while preserving context. On the other hand, Include-Exclude' can change the entire context of the analysis by filtering out specific subsets of the data or including additional data points \cite{law2019maqui}. 
This flexibility enables practitioners to tailor the data configuration to the specific demands of the analysis step. In addition, if practitioners have already started exploring certain strategies without considering the higher-level \intents, our framework can be useful for them to examine alternative \strategies \ that fulfill the same intent in different ways. \looseness -1

Finally, based on the input and desired output of each analysis step, practitioners can choose the \textit{specific} action and \textit{criteria} from the \techniques \ defined in our framework. The \textit{action-input-output-criteria} quartet provides a structured approach to selecting the most suitable technique for each analysis step.

% Based on the characteristics of the underlying data \cite{tukey1977exploratory}, our framework offers a comprehensive set of strategies to support these intents, for example, the developers may adopt segmenting if the sequences are long with varied events,  or aggregating if the type of events are few. Finally, based on the input and desired output of the analysis step, the practitioner can choose the action and criteria.

We plan to enhance the generative power of our framework by  developing a formal specification in future. This structured representation will enable automated generation of task-specific design recommendations and validation of event sequence analysis tool designs. 

% In the design stage, 
% Drawing upon the strategies and techniques, practitioners can determine how the identified analysis step can be supported. The framework offers a comprehensive set of options that can be applied to various event sequence analysis scenarios, allowing practitioners to choose from existing strategies and techniques or use them as a foundation for inventing new ones.

\vspace{-2mm}
\section{Limitations}
% \vspace{-2mm}

\rebuttal{While our framework provides a structured examination of event sequence visual analytic tasks, it is important to acknowledge certain limitations. }
% \rebuttal{First, the framework presented in this paper is derived from existing research, aiming to capture a broad spectrum of tasks supported by various systems. However, we do not claim this framework to be exhaustive in covering all possible strategies and techniques. As new research emerges, the framework will need to be revised and enhanced accordingly. 
% Future iterations may incorporate additional strategies and techniques that are not currently included. 
% % Therefore, while the framework serves as a valuable tool for evaluative purposes, it is imperative to recognize that it represents a snapshot of current knowledge and practices. 
% The taxonomy should be viewed as a living document that needs to be expanded and refined as new approaches and methods emerge.}
\rebuttal{First, the framework
% presented in this paper 
is derived from existing research and aims to capture a broad spectrum of tasks supported by various systems. However, we do not claim it to be exhaustive. As research advances, the framework will need to be revised and enhanced, incorporating additional strategies and techniques. The taxonomy should be viewed as a living document to be  expanded and refined as new methods emerge.
% as new approaches and methods emerge.
}\looseness -1

\rebuttal{Second, the boundaries between categories in any classification effort can be fuzzy and subject to multiple interpretations. For instance, 
at the \objectives \  level, 
% distinguishing between "Stage Progression" and "Pattern Exploration" exemplifies these challenges. 
``Stage Progression'' can be interpreted as a type of pattern, and may be subsumed under ``Pattern Exploration''. We decided to keep these two categories separate because ``Stage Progression'' involves segmenting sequences and tracking the progression \textit{across different segments}.
% , such as disease progression. 
In contrast, ``Pattern Exploration'' focuses on identifying recurring patterns \textit{across sequences} that may not involve segmentation.} 
% involves segmenting sequences and tracking the progression \textit{across different segments}, such as disease progression. 
% In contrast, "Pattern Exploration" focuses on identifying recurring patterns \textit{across sequences} that may not involve segmentation. In practice, such as in healthcare, analyzing patient treatment paths may involve both tracking individual stages and identifying overarching patterns, making rigid separation of these categories difficult.} \looseness -1
\rebuttal{At the \intents \  level,  categories such as ``Configure Data'' and ``Configure Visualization'' can overlap significantly. For example, zooming in on a visualization not only adjusts the visual representation but also acts as a filter on the underlying data. This overlap demonstrates the interconnected nature of data manipulation and visualization configuration, posing challenges in distinct categorization.}

\rebuttal{In the same light, the distinction between ``Simplify Data'' and ``Configure Data'' can be blurry, as simplification can be seen as a subset of configuration. We differentiate these two intents because data reduction is a crucial aspect of event sequence analytics \cite{zinat2023visual}. Strategies listed under ``Simplify Data'' focus on performing data reduction computations, while ``Configure Data'' is about dynamically choosing which aspects of data should be the focus of the current inspection.}

% configuration  are more trial and error or exploration-focused, whereas those under simplification reduce data components, such as total number of events, sequences or sequence length.}

    \rebuttal{The distinction between ``Aggregate'' and ``Summarize'' \strategies \ is another example of potential contention. ``Aggregate'' involves merging multiple events 
    % into representative entities, 
    without reducing the number of sequences or unique events. In contrast, ``Summarize'' focuses on extracting representative subsequences or patterns, potentially creating a lossy presentation by excluding rare events. The boundary between these \texttt{strategies} can be subtle, as one may argue that aggregation is one way to summarize.
    % particularly when both aggregation and summarization are applied iteratively or in combination.
    } 
    % \leo{I suggest removing this example, aggregation and segment are clearly different.}

    % The distinction between "Aggregate" and "Summarize" strategies is another area of contention. "Aggregate" involves combining multiple events into representative entities without reducing the number of sequences or unique events. In contrast, "Summarize" focuses on extracting representative subsequences or patterns, potentially creating a lossy presentation by excluding rare events. The boundary between these strategies can be subtle, especially when both are applied iteratively or in combination.

    \rebuttal{At the \techniques \ level,  the term ``Alignment'' is used in the literature to refer to both visual and data configuration. In our framework, alignment falls under the ``Augment Data'' intent,
    % we have placed alignment technique under the "Augment Data" intent, 
    focusing on data alignment operations such as establishing correspondences between events across multiple sequences based on a common reference point. This is conceptually similar to sequence alignment in fields like genomics \cite{li2010survey, apostolico1998sequence}. 
    % \leo{provide a reference.}
Nevertheless, alignment can also be a visual operation, involving repositioning of event sequences in a visualization (e.g., \cite{liu2016patternsandsequences,monroe2013eventflow}). To avoid confusion, we have used the term ``Reposition Sequences'' 
% within `Rearrange' strategy 
under ``Configure Visualization'' intent to describe such visual alignment actions. While this distinction helps maintain clarity within our framework, we acknowledge that our definition of the term may differ from its typical usage in visualization literature.} \looseness -2
% \leo{The discussion in the response letter is better than the text here.}

% At the technique level, the term "Alignment" is used in literature for both visual and data configuration. In our framework, alignment falls under the "Augment Data" intent, focusing on data alignment operations like establishing correspondences between events across multiple sequences, similar to sequence alignment in genomics. However, alignment can also involve repositioning event sequences in a visualization. To avoid confusion, we use "Reposition Sequences" under the "Configure Visualization" intent for visual alignment actions. This distinction maintains clarity, though our definition may differ from typical usage in visualization literature.

\rebuttal{These complexities underscore the evolving nature of our framework and the inherent challenges in categorizing diverse strategies and techniques. While we strive to provide a structured framework, 
% approach for comparative evaluations, 
some aspects 
% remain inherently nuanced and may 
defy clear-cut classification. }
% Future iterations of our framework will need to address these challenges by incorporating feedback and advancing methodologies that better accommodate interdisciplinary nuances and evolving research paradigms.}

% \rebuttal{These complexities underscore the evolving nature of our framework and the challenges in categorizing diverse strategies and techniques. While we aim to provide a structured approach for comparative evaluations, some aspects remain nuanced and resist clear-cut classification.}

\rebuttal{Third, our literature review was extensive but did not follow a formal systematic review process like PRISMA \cite{page2021prisma} in the healthcare domain, which ensures rigorous and reproducible examination. Future work could benefit from adopting such systematic approaches to minimize the risk of gaps and biases.} \looseness -2

% \rebuttal{Third, while we performed an extensive literature review, we did not follow a formal systematic review methodology, such as PRISMA \cite{page2021prisma}, which is commonly used in fields like healthcare to ensure a rigorous and reproducible examination of the literature. Future work could benefit from adopting such systematic approaches to minimize the risk of gaps and biases in the reviewed literature.}

% Finally, the current list of techniques, while derived from a comprehensive analysis of 58 papers, is not exhaustive and may appear somewhat scattered. As the authors indicate, the techniques can and should be extended based on new systems and case studies. Future work could focus on refining and organizing the techniques to provide a more coherent and complete set of building blocks for event sequence analysis tasks.

\rebuttal{Despite these limitations, we believe our framework takes an important step towards advancing the understanding of event sequence analysis tasks, setting the stage for future research and innovation.}
\section{Discussion and Future Work}

% \zinat{1. Importance: Theoretical Framework of Tasks in event sequence analysis: build conceptual understanding, then make it operationalizable. First step towards domain adaptation and transferring the learning from previous tools.
% }

% \zinat{Future Work: Build formal specification based on framework, construct analysis pipeline/ provenance, use specification to construct pipeline}

% \zinat{Future Work: Data property matters, include in framework to identify low level operations. Compute dataset properties and analyze What data operations make sense given data statistics? Automate operation flow for event sequence data analysis}

% \zinat{Future Work: Executable benchmark for tool evaluation from case studies}

\rebuttal{Our framework establishes a theoretical basis for characterizing a diverse range of tasks in event sequence analysis. 
% By providing a structured and comprehensive representation of tasks, our framework enables researchers and practitioners to build a conceptual understanding of the complex processes involved in event sequence analysis. 
Its hierarchical organization, from high-level \strategies \ to low-level \techniques, enables
% enhances operational feasibility, enabling practical application. The framework serves as an important step towards 
domain adaptation and knowledge transfer, by providing a common vocabulary for describing and comparing tasks across 
% different event sequence analysis 
applications.}

A promising direction for future work involves developing formal specifications based on our framework. Task specifications hold potential for constructing automated analysis pipelines and provenance.
% , enabling automation of workflows for event sequence data analysis.
% By leveraging the structured representation of our framework, researchers can design domain-agnostic and task-driven systems, that streamline analysis process and support reproducibility. 
Formal specifications can facilitate creation of intelligent assistance tools that suggest appropriate \techniques \ and effective analysis workflows based on analysis \objectives \ and dataset characteristics. \looseness -1

Another consideration is integrating data properties in the framework. Data characteristics, such as size, complexity, and heterogeneity, can impact both the choice of high-level intents and strategies as well as the performance of low-level operations. By analyzing the implications of data properties on these operations, researchers can develop more targeted and efficient analysis strategies. 
% Incorporating data properties into the framework can enable the automated recommendation of suitable techniques based on the statistical properties of the dataset, further enhancing the operationalizability of the framework.

% Furthermore, 
% the case studies and analysis workflows described using 
\rebuttal{These ideas for future work can pave the path towards 
% serve as a valuable resource for 
creating executable benchmarks for tool evaluation \cite{plaisant2004challenge}. By formalizing existing case studies along with associated \objectives \ and \techniques, researchers can develop standardized datasets and evaluation protocols.}
% that facilitate the comparison and assessment of different event sequence analysis tools. These benchmarks can provide insights into the strengths and weaknesses of existing tools, identify gaps in current approaches, and drive the development of more advanced and comprehensive event sequence analysis solutions.

\vspace{-2mm}
\section{Conclusion}
% \vspace{-2mm}
% \zinat{Emphasize the importance of our work again
% How it will help develop future tools and mappping of use case}

We present a multi-level framework to describe the \objectives, \intents, \strategies, and \techniques \  in event sequence visual analytics. Compared to existing event sequence task taxonomies, our framework enables more precise descriptions of tasks at multiple levels of granularity, and along multiple dimensions such as the \textit{action}, \textit{input}, \textit{output}, and \textit{criteria} associated with each technique.  The framework has the potential to promote knowledge transfer and generalization across domain-specific investigations, and lays the foundation for future research on formal specification languages and analysis strategy recommendations for event sequence data. 

\bibliographystyle{abbrv-doi-hyperref}
\bibliography{template}

\begin{thebibliography}{10}

\bibitem{ahn2013task}
J.-w. Ahn, C.~Plaisant, and B.~Shneiderman.
\newblock A task taxonomy for network evolution analysis.
\newblock {\em IEEE Transactions on Visualization and Computer Graphics}, 20(3):365--376, 2014. \href{https://doi.org/10.1109/TVCG.2013.238}
{doi: {{%
10\hspace{.1pt}\discretionary{.}{%
}{.}\hspace{.4pt}1109\discretionary{/}{%
}{/}TVCG\hspace{.1pt}\discretionary{.}{%
}{.}\hspace{.4pt}2013\hspace{.1pt}\discretionary{.}{%
}{.}\hspace{.4pt}238}}}


\bibitem{amar2005lowlevleanalytic}
R.~Amar, J.~Eagan, and J.~Stasko.
\newblock Low-level components of analytic activity in information visualization.
\newblock In {\em IEEE Symposium on Information Visualization, 2005. INFOVIS 2005.}, pp. 111--117, 2005. \href{https://doi.org/10.1109/INFVIS.2005.1532136}
{doi: {{%
10\hspace{.1pt}\discretionary{.}{%
}{.}\hspace{.4pt}1109\discretionary{/}{%
}{/}INFVIS\hspace{.1pt}\discretionary{.}{%
}{.}\hspace{.4pt}2005\hspace{.1pt}\discretionary{.}{%
}{.}\hspace{.4pt}1532136}}}


\bibitem{andrienko2006exploratory}
N.~Andrienko and G.~Andrienko.
\newblock {\em Exploratory Analysis of Spatial and Temporal Data: A Systematic Approach}.
\newblock Springer-Verlag, Berlin, 2005. \href{https://doi.org/10.1007/3-540-31190-4}
{doi: {{%
10\hspace{.1pt}\discretionary{.}{%
}{.}\hspace{.4pt}1007\discretionary{/}{%
}{/}3\discretionary{%
}{-}{-}540\discretionary{%
}{-}{-}31190\discretionary{%
}{-}{-}4}}}


\bibitem{apostolico1998sequence}
A.~APOSTOLICO and R.~GIANCARLO.
\newblock Sequence alignment in molecular biology.
\newblock {\em Journal of Computational Biology}, 5(2):173--196, 1998.
\newblock PMID: 9672827. \href{https://doi.org/10.1089/cmb.1998.5.173}
{doi: {{%
10\hspace{.1pt}\discretionary{.}{%
}{.}\hspace{.4pt}1089\discretionary{/}{%
}{/}cmb\hspace{.1pt}\discretionary{.}{%
}{.}\hspace{.4pt}1998\hspace{.1pt}\discretionary{.}{%
}{.}\hspace{.4pt}5\hspace{.1pt}\discretionary{.}{%
}{.}\hspace{.4pt}173}}}


\bibitem{di2020sequencebraiding}
S.~Bartolomeo, Y.~Zhang, F.~Sheng, and C.~Dunne.
\newblock Sequence braiding: Visual overviews of temporal event sequences and attributes.
\newblock {\em IEEE Transactions on Visualization and Computer Graphics}, 27(02):1353--1363, 2021. \href{https://doi.org/10.1109/TVCG.2020.3030442}
{doi: {{%
10\hspace{.1pt}\discretionary{.}{%
}{.}\hspace{.4pt}1109\discretionary{/}{%
}{/}TVCG\hspace{.1pt}\discretionary{.}{%
}{.}\hspace{.4pt}2020\hspace{.1pt}\discretionary{.}{%
}{.}\hspace{.4pt}3030442}}}


\bibitem{bernard2012guided}
J.~Bernard, T.~Ruppert, M.~Scherer, T.~Schreck, and J.~Kohlhammer.
\newblock Guided discovery of interesting relationships between time series clusters and metadata properties.
\newblock In {\em Proceedings of the 12th International Conference on Knowledge Management and Knowledge Technologies}. ACM, New York, 2012. \href{https://doi.org/10.1145/2362456.2362485}
{doi: {{%
10\hspace{.1pt}\discretionary{.}{%
}{.}\hspace{.4pt}1145\discretionary{/}{%
}{/}2362456\hspace{.1pt}\discretionary{.}{%
}{.}\hspace{.4pt}2362485}}}


\bibitem{brehmer2013typology}
M.~Brehmer and T.~Munzner.
\newblock A multi-level typology of abstract visualization tasks.
\newblock {\em IEEE Transactions on Visualization and Computer Graphics}, 19(12):2376--2385, 2013. \href{https://doi.org/10.1109/TVCG.2013.124}
{doi: {{%
10\hspace{.1pt}\discretionary{.}{%
}{.}\hspace{.4pt}1109\discretionary{/}{%
}{/}TVCG\hspace{.1pt}\discretionary{.}{%
}{.}\hspace{.4pt}2013\hspace{.1pt}\discretionary{.}{%
}{.}\hspace{.4pt}124}}}


\bibitem{cappers2018eventpad}
B.~C. Cappers, P.~N. Meessen, S.~Etalle, and J.~J. van Wijk.
\newblock Eventpad: Rapid malware analysis and reverse engineering using visual analytics.
\newblock In {\em 2018 IEEE Symposium on Visualization for Cyber Security (VizSec)}, pp. 1--8, 2018. \href{https://doi.org/10.1109/VIZSEC.2018.8709230}
{doi: {{%
10\hspace{.1pt}\discretionary{.}{%
}{.}\hspace{.4pt}1109\discretionary{/}{%
}{/}VIZSEC\hspace{.1pt}\discretionary{.}{%
}{.}\hspace{.4pt}2018\hspace{.1pt}\discretionary{.}{%
}{.}\hspace{.4pt}8709230}}}


\bibitem{cappers2018eventpad2}
B.~C. Cappers and J.~J. van Wijk.
\newblock Exploring multivariate event sequences using rules, aggregations, and selections.
\newblock {\em IEEE Transactions on Visualization and Computer Graphics}, 24(1):532--541, 2018. \href{https://doi.org/10.1109/TVCG.2017.2745278}
{doi: {{%
10\hspace{.1pt}\discretionary{.}{%
}{.}\hspace{.4pt}1109\discretionary{/}{%
}{/}TVCG\hspace{.1pt}\discretionary{.}{%
}{.}\hspace{.4pt}2017\hspace{.1pt}\discretionary{.}{%
}{.}\hspace{.4pt}2745278}}}


\bibitem{chen2017sequencesynopsis}
Y.~Chen, P.~Xu, and L.~Ren.
\newblock Sequence synopsis: Optimize visual summary of temporal event data.
\newblock {\em IEEE Transactions on Visualization and Computer Graphics}, 24(01):45--55, 2018. \href{https://doi.org/10.1109/TVCG.2017.2745083}
{doi: {{%
10\hspace{.1pt}\discretionary{.}{%
}{.}\hspace{.4pt}1109\discretionary{/}{%
}{/}TVCG\hspace{.1pt}\discretionary{.}{%
}{.}\hspace{.4pt}2017\hspace{.1pt}\discretionary{.}{%
}{.}\hspace{.4pt}2745083}}}


\bibitem{dextras2019segmentifier}
K.~Dextras-Romagnino and T.~Munzner.
\newblock Segmentifier: Interactive refinement of clickstream data.
\newblock {\em Computer Graphics Forum}, 38(3):623--634, 2019. \href{https://doi.org/10.1111/cgf.13715}
{doi: {{%
10\hspace{.1pt}\discretionary{.}{%
}{.}\hspace{.4pt}1111\discretionary{/}{%
}{/}cgf\hspace{.1pt}\discretionary{.}{%
}{.}\hspace{.4pt}13715}}}


\bibitem{du2019eventaction}
F.~Du, C.~Plaisant, N.~Spring, K.~Crowley, and B.~Shneiderman.
\newblock Eventaction: A visual analytics approach to explainable recommendation for event sequences.
\newblock {\em ACM Transactions on Interactive Intelligent Systems}, 9(4):1--31, 2019. \href{https://doi.org/10.1145/3301402}
{doi: {{%
10\hspace{.1pt}\discretionary{.}{%
}{.}\hspace{.4pt}1145\discretionary{/}{%
}{/}3301402}}}


\bibitem{du2016eventaction}
F.~Du, C.~Plaisant, N.~Spring, and B.~Shneiderman.
\newblock Eventaction: Visual analytics for temporal event sequence recommendation.
\newblock In {\em 2016 IEEE Conference on Visual Analytics Science and Technology (VAST)}, pp. 61--70, 2016. \href{https://doi.org/10.1109/VAST.2016.7883512}
{doi: {{%
10\hspace{.1pt}\discretionary{.}{%
}{.}\hspace{.4pt}1109\discretionary{/}{%
}{/}VAST\hspace{.1pt}\discretionary{.}{%
}{.}\hspace{.4pt}2016\hspace{.1pt}\discretionary{.}{%
}{.}\hspace{.4pt}7883512}}}


\bibitem{du2016volumeandvariety}
F.~Du, B.~Shneiderman, C.~Plaisant, S.~Malik, and A.~Perer.
\newblock Coping with volume and variety in temporal event sequences: Strategies for sharpening analytic focus.
\newblock {\em IEEE Transactions on Visualization and Computer Graphics}, 23(06):1636--1649, 2017. \href{https://doi.org/10.1109/TVCG.2016.2539960}
{doi: {{%
10\hspace{.1pt}\discretionary{.}{%
}{.}\hspace{.4pt}1109\discretionary{/}{%
}{/}TVCG\hspace{.1pt}\discretionary{.}{%
}{.}\hspace{.4pt}2016\hspace{.1pt}\discretionary{.}{%
}{.}\hspace{.4pt}2539960}}}


\bibitem{emmertstrreib2016emergent}
F.~Emmert-Streib, S.~Moutari, and M.~Dehmer.
\newblock The process of analyzing data is the emergent feature of data science.
\newblock {\em Frontiers in Genetics}, 7, 2016. \href{https://doi.org/10.3389/fgene.2016.00012}
{doi: {{%
10\hspace{.1pt}\discretionary{.}{%
}{.}\hspace{.4pt}3389\discretionary{/}{%
}{/}fgene\hspace{.1pt}\discretionary{.}{%
}{.}\hspace{.4pt}2016\hspace{.1pt}\discretionary{.}{%
}{.}\hspace{.4pt}00012}}}


\bibitem{gotz2013gapflow}
D.~Gotz, N.~Cao, E.~Goldbraich, and B.~Carmeli.
\newblock Gapflow : Visualizing gaps in care for medical treatment plans.
\newblock \url{https://gotz.web.unc.edu/research-project/gapflow/}, 2013.

\bibitem{gotz2014decisionflow}
D.~Gotz and H.~Stavropoulos.
\newblock Decisionflow: Visual analytics for high-dimensional temporal event sequence data.
\newblock {\em IEEE Transactions on Visualization and Computer Graphics}, 20(12):1783--1792, 2014. \href{https://doi.org/10.1109/TVCG.2014.2346682}
{doi: {{%
10\hspace{.1pt}\discretionary{.}{%
}{.}\hspace{.4pt}1109\discretionary{/}{%
}{/}TVCG\hspace{.1pt}\discretionary{.}{%
}{.}\hspace{.4pt}2014\hspace{.1pt}\discretionary{.}{%
}{.}\hspace{.4pt}2346682}}}


\bibitem{wongsuphasawat2012exploring}
D.~Gotz and K.~Wongsuphasawat.
\newblock Exploring flow, factors, and outcomes of temporal event sequences with the outflow visualization.
\newblock {\em IEEE Transactions on Visualization and Computer Graphics}, 18(12):2659--2668, 2012. \href{https://doi.org/10.1109/TVCG.2012.225}
{doi: {{%
10\hspace{.1pt}\discretionary{.}{%
}{.}\hspace{.4pt}1109\discretionary{/}{%
}{/}TVCG\hspace{.1pt}\discretionary{.}{%
}{.}\hspace{.4pt}2012\hspace{.1pt}\discretionary{.}{%
}{.}\hspace{.4pt}225}}}


\bibitem{gotz2019cadence}
D.~Gotz, J.~Zhang, W.~Wang, J.~Shrestha, and D.~Borland.
\newblock Visual analysis of high-dimensional event sequence data via dynamic hierarchical aggregation.
\newblock {\em IEEE Transactions on Visualization and Computer Graphics}, 26(01):440--450, 2020. \href{https://doi.org/10.1109/TVCG.2019.2934661}
{doi: {{%
10\hspace{.1pt}\discretionary{.}{%
}{.}\hspace{.4pt}1109\discretionary{/}{%
}{/}TVCG\hspace{.1pt}\discretionary{.}{%
}{.}\hspace{.4pt}2019\hspace{.1pt}\discretionary{.}{%
}{.}\hspace{.4pt}2934661}}}


\bibitem{gotz2008insightprovenance}
D.~Gotz and M.~X. Zhou.
\newblock Characterizing users' visual analytic activity for insight provenance.
\newblock {\em Information Visualization}, 8(1):42–55, 2009. \href{https://doi.org/10.1057/ivs.2008.31}
{doi: {{%
10\hspace{.1pt}\discretionary{.}{%
}{.}\hspace{.4pt}1057\discretionary{/}{%
}{/}ivs\hspace{.1pt}\discretionary{.}{%
}{.}\hspace{.4pt}2008\hspace{.1pt}\discretionary{.}{%
}{.}\hspace{.4pt}31}}}


\bibitem{guo2019uncertainty}
S.~Guo, F.~Du, S.~Malik, E.~Koh, S.~Kim, Z.~Liu, D.~Kim, H.~Zha, and N.~Cao.
\newblock Visualizing uncertainty and alternatives in event sequence predictions.
\newblock In {\em Proceedings of the 2019 CHI Conference on Human Factors in Computing Systems}, CHI '19, p. 1–12. ACM, New York, 2019. \href{https://doi.org/10.1145/3290605.3300803}
{doi: {{%
10\hspace{.1pt}\discretionary{.}{%
}{.}\hspace{.4pt}1145\discretionary{/}{%
}{/}3290605\hspace{.1pt}\discretionary{.}{%
}{.}\hspace{.4pt}3300803}}}


\bibitem{Guo2019anomaly}
S.~Guo, Z.~Jin, Q.~Chen, D.~Gotz, H.~Zha, and N.~Cao.
\newblock Visual anomaly detection in event sequence data.
\newblock In {\em 2019 IEEE International Conference on Big Data (Big Data)}. IEEE, 2019. \href{https://doi.org/10.1109/bigdata47090.2019.9005687}
{doi: {{%
10\hspace{.1pt}\discretionary{.}{%
}{.}\hspace{.4pt}1109\discretionary{/}{%
}{/}bigdata47090\hspace{.1pt}\discretionary{.}{%
}{.}\hspace{.4pt}2019\hspace{.1pt}\discretionary{.}{%
}{.}\hspace{.4pt}9005687}}}


\bibitem{guo2018eventthread2}
S.~Guo, Z.~Jin, D.~Gotz, F.~Du, H.~Zha, and N.~Cao.
\newblock Visual progression analysis of event sequence data.
\newblock {\em IEEE Transactions on Visualization and Computer Graphics}, 25:417--426, 2018. \href{https://doi.org/10.1109/TVCG.2018.2864885}
{doi: {{%
10\hspace{.1pt}\discretionary{.}{%
}{.}\hspace{.4pt}1109\discretionary{/}{%
}{/}TVCG\hspace{.1pt}\discretionary{.}{%
}{.}\hspace{.4pt}2018\hspace{.1pt}\discretionary{.}{%
}{.}\hspace{.4pt}2864885}}}


\bibitem{guo2017eventthread}
S.~Guo, K.~Xu, R.~Zhao, D.~Gotz, H.~Zha, and N.~Cao.
\newblock Eventthread: Visual summarization and stage analysis of event sequence data.
\newblock {\em IEEE Transactions on Visualization and Computer Graphics}, 24(1):56--65, 2018. \href{https://doi.org/10.1109/TVCG.2017.2745320}
{doi: {{%
10\hspace{.1pt}\discretionary{.}{%
}{.}\hspace{.4pt}1109\discretionary{/}{%
}{/}TVCG\hspace{.1pt}\discretionary{.}{%
}{.}\hspace{.4pt}2017\hspace{.1pt}\discretionary{.}{%
}{.}\hspace{.4pt}2745320}}}


\bibitem{guo2021survey}
Y.~Guo, S.~Guo, Z.~Jin, S.~Kaul, D.~Gotz, and N.~Cao.
\newblock Survey on visual analysis of event sequence data.
\newblock {\em IEEE Transactions on Visualization and Computer Graphics}, 28(12):5091--5112, 2022. \href{https://doi.org/10.1109/TVCG.2021.3100413}
{doi: {{%
10\hspace{.1pt}\discretionary{.}{%
}{.}\hspace{.4pt}1109\discretionary{/}{%
}{/}TVCG\hspace{.1pt}\discretionary{.}{%
}{.}\hspace{.4pt}2021\hspace{.1pt}\discretionary{.}{%
}{.}\hspace{.4pt}3100413}}}


\bibitem{tiphan2015tipovis}
Y.~Han, A.~Rozga, N.~Dimitrova, G.~D. Abowd, and J.~Stasko.
\newblock Visual analysis of proximal temporal relationships of social and communicative behaviors.
\newblock {\em Computer Graphics Forum}, 34(3):51--60, 2015. \href{https://doi.org/10.1111/cgf.12617}
{doi: {{%
10\hspace{.1pt}\discretionary{.}{%
}{.}\hspace{.4pt}1111\discretionary{/}{%
}{/}cgf\hspace{.1pt}\discretionary{.}{%
}{.}\hspace{.4pt}12617}}}


\bibitem{heer2012interactive}
J.~Heer and B.~Shneiderman.
\newblock Interactive dynamics for visual analysis: A taxonomy of tools that support the fluent and flexible use of visualizations.
\newblock {\em Queue}, 10(2):30–55, 2012. \href{https://doi.org/10.1145/2133416.2146416}
{doi: {{%
10\hspace{.1pt}\discretionary{.}{%
}{.}\hspace{.4pt}1145\discretionary{/}{%
}{/}2133416\hspace{.1pt}\discretionary{.}{%
}{.}\hspace{.4pt}2146416}}}


\bibitem{hu2016sententree}
M.~Hu, K.~Wongsuphasawat, and J.~Stasko.
\newblock Visualizing social media content with sententree.
\newblock {\em IEEE transactions on visualization and computer graphics}, 23(1):621--630, 2016.

\bibitem{jentner2019techniquesforpatterns}
W.~Jentner and D.~A. Keim.
\newblock {\em Visualization and visual analytic techniques for patterns}.
\newblock Springer, 2019.

\bibitem{jin2021seqcausal}
Z.~Jin, S.~Guo, N.~Chen, D.~Weiskopf, D.~Gotz, and N.~Cao.
\newblock Visual causality analysis of event sequence data.
\newblock {\em IEEE Transactions on Visualization and Computer Graphics}, 27(2):1343--1352, 2021. \href{https://doi.org/10.1109/TVCG.2020.3030465}
{doi: {{%
10\hspace{.1pt}\discretionary{.}{%
}{.}\hspace{.4pt}1109\discretionary{/}{%
}{/}TVCG\hspace{.1pt}\discretionary{.}{%
}{.}\hspace{.4pt}2020\hspace{.1pt}\discretionary{.}{%
}{.}\hspace{.4pt}3030465}}}


\bibitem{j02014livegantt}
J.~Jo, J.~Huh, J.~Park, B.~Kim, and J.~Seo.
\newblock Livegantt: Interactively visualizing a large manufacturing schedule.
\newblock {\em IEEE Transactions on Visualization and Computer Graphics}, 20(12):2329--2338, 2014. \href{https://doi.org/10.1109/TVCG.2014.2346454}
{doi: {{%
10\hspace{.1pt}\discretionary{.}{%
}{.}\hspace{.4pt}1109\discretionary{/}{%
}{/}TVCG\hspace{.1pt}\discretionary{.}{%
}{.}\hspace{.4pt}2014\hspace{.1pt}\discretionary{.}{%
}{.}\hspace{.4pt}2346454}}}


\bibitem{kim2020seq2vec}
H.~J. Kim, S.~E. Hong, and K.~J. Cha.
\newblock seq2vec: Analyzing sequential data using multi-rank embedding vectors.
\newblock {\em Electronic Commerce Research and Applications}, 43:101003, 2020. \href{https://doi.org/10.1016/j.elerap.2020.101003}
{doi: {{%
10\hspace{.1pt}\discretionary{.}{%
}{.}\hspace{.4pt}1016\discretionary{/}{%
}{/}j\hspace{.1pt}\discretionary{.}{%
}{.}\hspace{.4pt}elerap\hspace{.1pt}\discretionary{.}{%
}{.}\hspace{.4pt}2020\hspace{.1pt}\discretionary{.}{%
}{.}\hspace{.4pt}101003}}}


\bibitem{krause2015supporting}
J.~Krause, A.~Perer, and H.~Stavropoulos.
\newblock Supporting iterative cohort construction with visual temporal queries.
\newblock {\em IEEE Transactions on Visualization and Computer Graphics}, 22(01):91--100, 2016. \href{https://doi.org/10.1109/TVCG.2015.2467622}
{doi: {{%
10\hspace{.1pt}\discretionary{.}{%
}{.}\hspace{.4pt}1109\discretionary{/}{%
}{/}TVCG\hspace{.1pt}\discretionary{.}{%
}{.}\hspace{.4pt}2015\hspace{.1pt}\discretionary{.}{%
}{.}\hspace{.4pt}2467622}}}


\bibitem{Kwon2018RetainVisVA}
B.~Kwon, M.~Choi, J.~Kim, E.~Choi, Y.~Kim, S.~Kwon, J.~Sun, and J.~Choo.
\newblock Retainvis: Visual analytics with interpretable and interactive recurrent neural networks on electronic medical records.
\newblock {\em IEEE Transactions on Visualization and Computer Graphics}, 25(01):299--309, 2019. \href{https://doi.org/10.1109/TVCG.2018.2865027}
{doi: {{%
10\hspace{.1pt}\discretionary{.}{%
}{.}\hspace{.4pt}1109\discretionary{/}{%
}{/}TVCG\hspace{.1pt}\discretionary{.}{%
}{.}\hspace{.4pt}2018\hspace{.1pt}\discretionary{.}{%
}{.}\hspace{.4pt}2865027}}}


\bibitem{kwon2020dpvis}
B.~C. Kwon, V.~Anand, K.~A. Severson, S.~Ghosh, Z.~Sun, B.~I. Frohnert, M.~Lundgren, and K.~Ng.
\newblock Dpvis: Visual analytics with hidden markov models for disease progression pathways.
\newblock {\em IEEE Transactions on Visualization and Computer Graphics}, 27(9):3685--3700, 2021. \href{https://doi.org/10.1109/TVCG.2020.2985689}
{doi: {{%
10\hspace{.1pt}\discretionary{.}{%
}{.}\hspace{.4pt}1109\discretionary{/}{%
}{/}TVCG\hspace{.1pt}\discretionary{.}{%
}{.}\hspace{.4pt}2020\hspace{.1pt}\discretionary{.}{%
}{.}\hspace{.4pt}2985689}}}


\bibitem{lam2007sessionviewer}
H.~Lam, D.~Russell, D.~Tang, and T.~Munzner.
\newblock Session viewer: Visual exploratory analysis of web session logs.
\newblock In {\em 2007 IEEE Symposium on Visual Analytics Science and Technology}, pp. 147--154, 2007. \href{https://doi.org/10.1109/VAST.2007.4389008}
{doi: {{%
10\hspace{.1pt}\discretionary{.}{%
}{.}\hspace{.4pt}1109\discretionary{/}{%
}{/}VAST\hspace{.1pt}\discretionary{.}{%
}{.}\hspace{.4pt}2007\hspace{.1pt}\discretionary{.}{%
}{.}\hspace{.4pt}4389008}}}


\bibitem{lam2017goalstotasks}
H.~Lam, M.~Tory, and T.~Munzner.
\newblock Bridging from goals to tasks with design study analysis reports.
\newblock {\em IEEE Transactions on Visualization and Computer Graphics}, 24(1):435--445, 2018. \href{https://doi.org/10.1109/TVCG.2017.2744319}
{doi: {{%
10\hspace{.1pt}\discretionary{.}{%
}{.}\hspace{.4pt}1109\discretionary{/}{%
}{/}TVCG\hspace{.1pt}\discretionary{.}{%
}{.}\hspace{.4pt}2017\hspace{.1pt}\discretionary{.}{%
}{.}\hspace{.4pt}2744319}}}


\bibitem{law2019maqui}
P.-M. Law, Z.~Liu, S.~Malik, and R.~C. Basole.
\newblock Maqui: Interweaving queries and pattern mining for recursive event sequence exploration.
\newblock {\em IEEE Transactions on Visualization and Computer Graphics}, 25(1):396--406, 2019. \href{https://doi.org/10.1109/TVCG.2018.2864886}
{doi: {{%
10\hspace{.1pt}\discretionary{.}{%
}{.}\hspace{.4pt}1109\discretionary{/}{%
}{/}TVCG\hspace{.1pt}\discretionary{.}{%
}{.}\hspace{.4pt}2018\hspace{.1pt}\discretionary{.}{%
}{.}\hspace{.4pt}2864886}}}


\bibitem{lekschas2020peax}
F.~Lekschas, B.~Peterson, D.~Haehn, E.~Ma, N.~Gehlenborg, and H.~Pfister.
\newblock Peax: Interactive visual pattern search in sequential data using unsupervised deep representation learning.
\newblock {\em Computer Graphics Forum}, 39(3):167--179, 2020. \href{https://doi.org/10.1111/cgf.13971}
{doi: {{%
10\hspace{.1pt}\discretionary{.}{%
}{.}\hspace{.4pt}1111\discretionary{/}{%
}{/}cgf\hspace{.1pt}\discretionary{.}{%
}{.}\hspace{.4pt}13971}}}


\bibitem{li2010survey}
H.~Li and N.~Homer.
\newblock {A survey of sequence alignment algorithms for next-generation sequencing}.
\newblock {\em Briefings in Bioinformatics}, 11(5):473--483, 2010. \href{https://doi.org/10.1093/bib/bbq015}
{doi: {{%
10\hspace{.1pt}\discretionary{.}{%
}{.}\hspace{.4pt}1093\discretionary{/}{%
}{/}bib\discretionary{/}{%
}{/}bbq015}}}


\bibitem{linden2023flex}
S.~v.~d. Linden, B.~M. Wulterkens, M.~M.~v. Gilst, S.~Overeem, C.~v. Pul, A.~Vilanova, and S.~v.~d. Elzen.
\newblock {FlexEvent: going beyond Case-Centric Exploration and Analysis of Multivariate Event Sequences}.
\newblock {\em Computer Graphics Forum}, 2023. \href{https://doi.org/10.1111/cgf.14820}
{doi: {{%
10\hspace{.1pt}\discretionary{.}{%
}{.}\hspace{.4pt}1111\discretionary{/}{%
}{/}cgf\hspace{.1pt}\discretionary{.}{%
}{.}\hspace{.4pt}14820}}}


\bibitem{liu2017coreflow}
Z.~Liu, B.~Kerr, M.~Dontcheva, J.~Grover, M.~Hoffman, and A.~Wilson.
\newblock Coreflow: Extracting and visualizing branching patterns from event sequences.
\newblock {\em Computer Graphics Forum}, 36(3):527--538, 2017. \href{https://doi.org/10.1111/cgf.13208}
{doi: {{%
10\hspace{.1pt}\discretionary{.}{%
}{.}\hspace{.4pt}1111\discretionary{/}{%
}{/}cgf\hspace{.1pt}\discretionary{.}{%
}{.}\hspace{.4pt}13208}}}


\bibitem{liu2016patternsandsequences}
Z.~Liu, Y.~Wang, M.~Dontcheva, M.~Hoffman, S.~Walker, and A.~Wilson.
\newblock Patterns and sequences: Interactive exploration of clickstreams to understand common visitor paths.
\newblock {\em IEEE Transactions on Visualization and Computer Graphics}, 23(1):321--330, 2017. \href{https://doi.org/10.1109/TVCG.2016.2598797}
{doi: {{%
10\hspace{.1pt}\discretionary{.}{%
}{.}\hspace{.4pt}1109\discretionary{/}{%
}{/}TVCG\hspace{.1pt}\discretionary{.}{%
}{.}\hspace{.4pt}2016\hspace{.1pt}\discretionary{.}{%
}{.}\hspace{.4pt}2598797}}}


\bibitem{magallanes2021sequen}
J.~Magallanes, T.~Stone, P.~D. Morris, S.~Mason, S.~Wood, and M.-C. Villa-Uriol.
\newblock Sequen-c: A multilevel overview of temporal event sequences.
\newblock {\em IEEE Transactions on Visualization and Computer Graphics}, 28(1):901--911, 2022. \href{https://doi.org/10.1109/TVCG.2021.3114868}
{doi: {{%
10\hspace{.1pt}\discretionary{.}{%
}{.}\hspace{.4pt}1109\discretionary{/}{%
}{/}TVCG\hspace{.1pt}\discretionary{.}{%
}{.}\hspace{.4pt}2021\hspace{.1pt}\discretionary{.}{%
}{.}\hspace{.4pt}3114868}}}


\bibitem{malik2015coco}
S.~Malik, F.~Du, M.~Monroe, E.~Onukwugha, C.~Plaisant, and B.~Shneiderman.
\newblock Cohort comparison of event sequences with balanced integration of visual analytics and statistics.
\newblock In {\em Proceedings of the 20th International Conference on Intelligent User Interfaces}, IUI '15, p. 38–49. ACM, New York, 2015. \href{https://doi.org/10.1145/2678025.2701407}
{doi: {{%
10\hspace{.1pt}\discretionary{.}{%
}{.}\hspace{.4pt}1145\discretionary{/}{%
}{/}2678025\hspace{.1pt}\discretionary{.}{%
}{.}\hspace{.4pt}2701407}}}


\bibitem{ming2019protosteer}
Y.~Ming, P.~Xu, F.~Cheng, H.~Qu, and L.~Ren.
\newblock Protosteer: Steering deep sequence model with prototypes.
\newblock {\em IEEE Transactions on Visualization and Computer Graphics}, 26(1):238--248, 2020. \href{https://doi.org/10.1109/TVCG.2019.2934267}
{doi: {{%
10\hspace{.1pt}\discretionary{.}{%
}{.}\hspace{.4pt}1109\discretionary{/}{%
}{/}TVCG\hspace{.1pt}\discretionary{.}{%
}{.}\hspace{.4pt}2019\hspace{.1pt}\discretionary{.}{%
}{.}\hspace{.4pt}2934267}}}


\bibitem{monroe2013eventflow}
M.~Monroe, R.~Lan, H.~Lee, C.~Plaisant, and B.~Shneiderman.
\newblock Temporal event sequence simplification.
\newblock {\em IEEE Transactions on Visualization and Computer Graphics}, 19(12):2227--2236, 2013. \href{https://doi.org/10.1109/TVCG.2013.200}
{doi: {{%
10\hspace{.1pt}\discretionary{.}{%
}{.}\hspace{.4pt}1109\discretionary{/}{%
}{/}TVCG\hspace{.1pt}\discretionary{.}{%
}{.}\hspace{.4pt}2013\hspace{.1pt}\discretionary{.}{%
}{.}\hspace{.4pt}200}}}


\bibitem{page2021prisma}
M.~J. Page, J.~E. McKenzie, P.~M. Bossuyt, I.~Boutron, T.~C. Hoffmann, C.~D. Mulrow, L.~Shamseer, J.~M. Tetzlaff, E.~A. Akl, S.~E. Brennan, R.~Chou, J.~Glanville, J.~M. Grimshaw, A.~Hr{\'o}bjartsson, M.~M. Lalu, T.~Li, E.~W. Loder, E.~Mayo-Wilson, S.~McDonald, L.~A. McGuinness, L.~A. Stewart, J.~Thomas, A.~C. Tricco, V.~A. Welch, P.~Whiting, and D.~Moher.
\newblock The prisma 2020 statement: an updated guideline for reporting systematic reviews.
\newblock {\em BMJ}, 372, 2021. \href{https://doi.org/10.1136/bmj.n71}
{doi: {{%
10\hspace{.1pt}\discretionary{.}{%
}{.}\hspace{.4pt}1136\discretionary{/}{%
}{/}bmj\hspace{.1pt}\discretionary{.}{%
}{.}\hspace{.4pt}n71}}}


\bibitem{peiris2022data}
Y.~Peiris, C.~Barth, E.~M. Huang, and J.~Bernard.
\newblock A data-centric methodology and task typology for time-stamped event sequences.
\newblock In {\em 2022 IEEE Evaluation and Beyond - Methodological Approaches for Visualization (BELIV)}, pp. 66--76. IEEE Computer Society, Los Alamitos, 2022. \href{https://doi.org/10.1109/BELIV57783.2022.00012}
{doi: {{%
10\hspace{.1pt}\discretionary{.}{%
}{.}\hspace{.4pt}1109\discretionary{/}{%
}{/}BELIV57783\hspace{.1pt}\discretionary{.}{%
}{.}\hspace{.4pt}2022\hspace{.1pt}\discretionary{.}{%
}{.}\hspace{.4pt}00012}}}


\bibitem{perer2013careflow}
A.~Perer and D.~Gotz.
\newblock Data-driven exploration of care plans for patients.
\newblock In {\em CHI '13 Extended Abstracts on Human Factors in Computing Systems}, CHI EA '13, p. 439–444. ACM, New York, 2013. \href{https://doi.org/10.1145/2468356.2468434}
{doi: {{%
10\hspace{.1pt}\discretionary{.}{%
}{.}\hspace{.4pt}1145\discretionary{/}{%
}{/}2468356\hspace{.1pt}\discretionary{.}{%
}{.}\hspace{.4pt}2468434}}}


\bibitem{perer2014frequence}
A.~Perer and F.~Wang.
\newblock Frequence: interactive mining and visualization of temporal frequent event sequences.
\newblock In {\em Proceedings of the 19th International Conference on Intelligent User Interfaces}, IUI '14, p. 153–162. ACM, New York, 2014. \href{https://doi.org/10.1145/2557500.2557508}
{doi: {{%
10\hspace{.1pt}\discretionary{.}{%
}{.}\hspace{.4pt}1145\discretionary{/}{%
}{/}2557500\hspace{.1pt}\discretionary{.}{%
}{.}\hspace{.4pt}2557508}}}


\bibitem{plaisant2004challenge}
C.~Plaisant.
\newblock The challenge of information visualization evaluation.
\newblock In {\em Proceedings of the Working Conference on Advanced Visual Interfaces}, AVI '04, p. 109–116. ACM, New York, 2004. \href{https://doi.org/10.1145/989863.989880}
{doi: {{%
10\hspace{.1pt}\discretionary{.}{%
}{.}\hspace{.4pt}1145\discretionary{/}{%
}{/}989863\hspace{.1pt}\discretionary{.}{%
}{.}\hspace{.4pt}989880}}}


\bibitem{plaisant1996lifelines}
C.~Plaisant, B.~Milash, A.~Rose, S.~Widoff, and B.~Shneiderman.
\newblock Lifelines: visualizing personal histories.
\newblock In {\em Proceedings of the SIGCHI Conference on Human Factors in Computing Systems}, CHI '96, p. 221–227. ACM, New York, 1996. \href{https://doi.org/10.1145/238386.238493}
{doi: {{%
10\hspace{.1pt}\discretionary{.}{%
}{.}\hspace{.4pt}1145\discretionary{/}{%
}{/}238386\hspace{.1pt}\discretionary{.}{%
}{.}\hspace{.4pt}238493}}}


\bibitem{plaisant2016diversity}
C.~Plaisant and B.~Shneiderman.
\newblock The diversity of data and tasks in event analytics.
\newblock \url{https://eventevent.github.io/papers/EVENT_2016_paper_13.pdf}, 2016.

\bibitem{qi2020stbins}
J.~Qi, V.~Bloemen, S.~Wang, J.~van Wijk, and H.~van~de Wetering.
\newblock Stbins: Visual tracking and comparison of multiple data sequences using temporal binning.
\newblock {\em IEEE Transactions on Visualization and Computer Graphics}, 26(1):1054--1063, 2020. \href{https://doi.org/10.1109/TVCG.2019.2934289}
{doi: {{%
10\hspace{.1pt}\discretionary{.}{%
}{.}\hspace{.4pt}1109\discretionary{/}{%
}{/}TVCG\hspace{.1pt}\discretionary{.}{%
}{.}\hspace{.4pt}2019\hspace{.1pt}\discretionary{.}{%
}{.}\hspace{.4pt}2934289}}}


\bibitem{rind_task_2016}
A.~Rind, W.~Aigner, M.~Wagner, S.~Miksch, and T.~Lammarsch.
\newblock Task cube: A three-dimensional conceptual space of user tasks in visualization design and evaluation.
\newblock {\em Information Visualization}, 15(4):288--300, 2016. \href{https://doi.org/10.1177/1473871615621602}
{doi: {{%
10\hspace{.1pt}\discretionary{.}{%
}{.}\hspace{.4pt}1177\discretionary{/}{%
}{/}1473871615621602}}}


\bibitem{schulz2013design}
H.-J. Schulz, T.~Nocke, M.~Heitzler, and H.~Schumann.
\newblock A design space of visualization tasks.
\newblock {\em IEEE Transactions on Visualization and Computer Graphics}, 19(12):2366--2375, 2013. \href{https://doi.org/10.1109/TVCG.2013.120}
{doi: {{%
10\hspace{.1pt}\discretionary{.}{%
}{.}\hspace{.4pt}1109\discretionary{/}{%
}{/}TVCG\hspace{.1pt}\discretionary{.}{%
}{.}\hspace{.4pt}2013\hspace{.1pt}\discretionary{.}{%
}{.}\hspace{.4pt}120}}}


\bibitem{sedlmair2012design}
M.~Sedlmair, M.~Meyer, and T.~Munzner.
\newblock Design study methodology: Reflections from the trenches and the stacks.
\newblock {\em IEEE Transactions on Visualization and Computer Graphics}, 18(12):2431--2440, 2012. \href{https://doi.org/10.1109/TVCG.2012.213}
{doi: {{%
10\hspace{.1pt}\discretionary{.}{%
}{.}\hspace{.4pt}1109\discretionary{/}{%
}{/}TVCG\hspace{.1pt}\discretionary{.}{%
}{.}\hspace{.4pt}2012\hspace{.1pt}\discretionary{.}{%
}{.}\hspace{.4pt}213}}}


\bibitem{shneiderman2003eyes}
B.~Shneiderman.
\newblock The eyes have it: a task by data type taxonomy for information visualizations.
\newblock In {\em Proceedings 1996 IEEE Symposium on Visual Languages}, pp. 336--343, 1996. \href{https://doi.org/10.1109/VL.1996.545307}
{doi: {{%
10\hspace{.1pt}\discretionary{.}{%
}{.}\hspace{.4pt}1109\discretionary{/}{%
}{/}VL\hspace{.1pt}\discretionary{.}{%
}{.}\hspace{.4pt}1996\hspace{.1pt}\discretionary{.}{%
}{.}\hspace{.4pt}545307}}}


\bibitem{shneiderman2016eventquartet}
B.~Shneiderman.
\newblock The event quartet: How visual analytics works for temporal data.
\newblock \url{https://eventevent.github.io/papers/EVENT_2016_paper_7.pdf}, 2016.

\bibitem{valiati2006taxonomy}
E.~R.~A. Valiati, M.~S. Pimenta, and C.~M. D.~S. Freitas.
\newblock A taxonomy of tasks for guiding the evaluation of multidimensional visualizations.
\newblock In {\em Proceedings of the 2006 AVI Workshop on BEyond Time and Errors: Novel Evaluation Methods for Information Visualization}, BELIV '06, p. 1–6. ACM, New York, 2006. \href{https://doi.org/10.1145/1168149.1168169}
{doi: {{%
10\hspace{.1pt}\discretionary{.}{%
}{.}\hspace{.4pt}1145\discretionary{/}{%
}{/}1168149\hspace{.1pt}\discretionary{.}{%
}{.}\hspace{.4pt}1168169}}}


\bibitem{vrotsou2009activitree}
K.~Vrotsou, J.~Johansson, and M.~Cooper.
\newblock Activitree: Interactive visual exploration of sequences in event-based data using graph similarity.
\newblock {\em IEEE Transactions on Visualization and Computer Graphics}, 15(6):945--952, 2009. \href{https://doi.org/10.1109/TVCG.2009.117}
{doi: {{%
10\hspace{.1pt}\discretionary{.}{%
}{.}\hspace{.4pt}1109\discretionary{/}{%
}{/}TVCG\hspace{.1pt}\discretionary{.}{%
}{.}\hspace{.4pt}2009\hspace{.1pt}\discretionary{.}{%
}{.}\hspace{.4pt}117}}}


\bibitem{vrotsou2018eloquence}
K.~Vrotsou and A.~Nordman.
\newblock Exploratory visual sequence mining based on pattern-growth.
\newblock {\em IEEE Transactions on Visualization and Computer Graphics}, 25(8):2597--2610, 2019. \href{https://doi.org/10.1109/TVCG.2018.2848247}
{doi: {{%
10\hspace{.1pt}\discretionary{.}{%
}{.}\hspace{.4pt}1109\discretionary{/}{%
}{/}TVCG\hspace{.1pt}\discretionary{.}{%
}{.}\hspace{.4pt}2018\hspace{.1pt}\discretionary{.}{%
}{.}\hspace{.4pt}2848247}}}


\bibitem{wang2008aligning}
T.~D. Wang, C.~Plaisant, A.~J. Quinn, R.~Stanchak, S.~Murphy, and B.~Shneiderman.
\newblock Aligning temporal data by sentinel events: discovering patterns in electronic health records.
\newblock In {\em Proceedings of the SIGCHI Conference on Human Factors in Computing Systems}, CHI '08, p. 457–466. ACM, New York, 2008. \href{https://doi.org/10.1145/1357054.1357129}
{doi: {{%
10\hspace{.1pt}\discretionary{.}{%
}{.}\hspace{.4pt}1145\discretionary{/}{%
}{/}1357054\hspace{.1pt}\discretionary{.}{%
}{.}\hspace{.4pt}1357129}}}


\bibitem{wang2009temporal}
T.~D. Wang, C.~Plaisant, B.~Shneiderman, N.~Spring, D.~Roseman, G.~Marchand, V.~Mukherjee, and M.~Smith.
\newblock Temporal summaries: Supporting temporal categorical searching, aggregation and comparison.
\newblock {\em IEEE Transactions on Visualization and Computer Graphics}, 15(6):1049--1056, 2009. \href{https://doi.org/10.1109/TVCG.2009.187}
{doi: {{%
10\hspace{.1pt}\discretionary{.}{%
}{.}\hspace{.4pt}1109\discretionary{/}{%
}{/}TVCG\hspace{.1pt}\discretionary{.}{%
}{.}\hspace{.4pt}2009\hspace{.1pt}\discretionary{.}{%
}{.}\hspace{.4pt}187}}}


\bibitem{Wongsuphasawat2011Outflow}
K.~Wongsuphasawat and D.~Gotz.
\newblock Outflow : Visualizing patient flow by symptoms and outcome.
\newblock \url{https://gotz.web.unc.edu/wp-content/uploads/sites/5664/2013/10/wongsuphasawat_ieee_visweek_vahc_2011.pdf}, 2011.

\bibitem{wongsuphasawat2011lifeflow}
K.~Wongsuphasawat, J.~A. Guerra~G\'{o}mez, C.~Plaisant, T.~D. Wang, M.~Taieb-Maimon, and B.~Shneiderman.
\newblock Lifeflow: visualizing an overview of event sequences.
\newblock In {\em Proceedings of the SIGCHI Conference on Human Factors in Computing Systems}, CHI '11, p. 1747–1756. ACM, New York, 2011. \href{https://doi.org/10.1145/1978942.1979196}
{doi: {{%
10\hspace{.1pt}\discretionary{.}{%
}{.}\hspace{.4pt}1145\discretionary{/}{%
}{/}1978942\hspace{.1pt}\discretionary{.}{%
}{.}\hspace{.4pt}1979196}}}


\bibitem{wu2022rasipam}
J.~Wu, D.~Liu, Z.~Guo, and Y.~Wu.
\newblock Rasipam: Interactive pattern mining of multivariate event sequences in racket sports.
\newblock {\em IEEE Transactions on Visualization and Computer Graphics}, 29(1):940--950, 2023. \href{https://doi.org/10.1109/TVCG.2022.3209452}
{doi: {{%
10\hspace{.1pt}\discretionary{.}{%
}{.}\hspace{.4pt}1109\discretionary{/}{%
}{/}TVCG\hspace{.1pt}\discretionary{.}{%
}{.}\hspace{.4pt}2022\hspace{.1pt}\discretionary{.}{%
}{.}\hspace{.4pt}3209452}}}


\bibitem{wu2014opinionflow}
Y.~Wu, S.~Liu, K.~Yan, M.~Liu, and F.~Wu.
\newblock Opinionflow: Visual analysis of opinion diffusion on social media.
\newblock {\em IEEE Transactions on Visualization and Computer Graphics}, 20(12):1763--1772, 2014. \href{https://doi.org/10.1109/TVCG.2014.2346920}
{doi: {{%
10\hspace{.1pt}\discretionary{.}{%
}{.}\hspace{.4pt}1109\discretionary{/}{%
}{/}TVCG\hspace{.1pt}\discretionary{.}{%
}{.}\hspace{.4pt}2014\hspace{.1pt}\discretionary{.}{%
}{.}\hspace{.4pt}2346920}}}


\bibitem{yi2007toward}
J.~S. Yi, Y.~a. Kang, J.~Stasko, and J.~Jacko.
\newblock Toward a deeper understanding of the role of interaction in information visualization.
\newblock {\em IEEE Transactions on Visualization and Computer Graphics}, 13(6):1224--1231, 2007. \href{https://doi.org/10.1109/TVCG.2007.70515}
{doi: {{%
10\hspace{.1pt}\discretionary{.}{%
}{.}\hspace{.4pt}1109\discretionary{/}{%
}{/}TVCG\hspace{.1pt}\discretionary{.}{%
}{.}\hspace{.4pt}2007\hspace{.1pt}\discretionary{.}{%
}{.}\hspace{.4pt}70515}}}


\bibitem{zgraggen2015squeries}
E.~Zgraggen, S.~M. Drucker, D.~Fisher, and R.~DeLine.
\newblock (s|qu)eries: Visual regular expressions for querying and exploring event sequences.
\newblock In {\em Proceedings of the 33rd Annual ACM Conference on Human Factors in Computing Systems}, CHI '15, p. 2683–2692. ACM, New York, 2015. \href{https://doi.org/10.1145/2702123.2702262}
{doi: {{%
10\hspace{.1pt}\discretionary{.}{%
}{.}\hspace{.4pt}1145\discretionary{/}{%
}{/}2702123\hspace{.1pt}\discretionary{.}{%
}{.}\hspace{.4pt}2702262}}}


\bibitem{zhao2015matrixwave}
J.~Zhao, Z.~Liu, M.~Dontcheva, A.~Hertzmann, and A.~Wilson.
\newblock Matrixwave: Visual comparison of event sequence data.
\newblock In {\em Proceedings of the 33rd Annual ACM Conference on Human Factors in Computing Systems}, CHI '15, p. 259–268. ACM, New York, 2015. \href{https://doi.org/10.1145/2702123.2702419}
{doi: {{%
10\hspace{.1pt}\discretionary{.}{%
}{.}\hspace{.4pt}1145\discretionary{/}{%
}{/}2702123\hspace{.1pt}\discretionary{.}{%
}{.}\hspace{.4pt}2702419}}}


\bibitem{zinat2023visual}
K.~T. Zinat, J.~Yang, A.~Gandhi, N.~Mitra, and Z.~Liu.
\newblock A comparative evaluation of visual summarization techniques for event sequences.
\newblock {\em Computer Graphics Forum}, 42(3):173--185, 2023. \href{https://doi.org/10.1111/cgf.14821}
{doi: {{%
10\hspace{.1pt}\discretionary{.}{%
}{.}\hspace{.4pt}1111\discretionary{/}{%
}{/}cgf\hspace{.1pt}\discretionary{.}{%
}{.}\hspace{.4pt}14821}}}


\end{thebibliography}
\end{document}